\documentclass[a4paper,11pt]{article}
\usepackage{graphicx}
\pdfoutput=1 

\usepackage{jcappub} 
\usepackage{mathrsfs}
\usepackage[T1]{fontenc} 
\usepackage{bbold}
\usepackage{url}
\usepackage{hyperref}
\usepackage[english]{babel}
\usepackage[utf8]{inputenc}
\usepackage{amsmath}
\usepackage{color}
\usepackage{amsfonts}
\usepackage{amssymb}
\usepackage{mathtools}
\usepackage{placeins}
\usepackage{float}
\usepackage{tabularx}
\usepackage{adjustbox}
\usepackage{caption}
\usepackage{subcaption}

\begin{document}
\title{\boldmath Can breakdown of perturbation in the $\alpha$-attractor inflation lead to PBH formation?}
\author[a]{Chitrak Sarkar}
\author[a]{Arunoday Sarkar}
\author[b]{Basundhara Ghosh}
\author[a]{Buddhadeb Ghosh}

\affiliation[a]{Centre of Advanced Studies, Department of Physics, The University of Burdwan,\\Burdwan 713 104, India}
\affiliation[b]{Department of Physics, Indian Institute of Science, Bangalore, Karnataka, 560012, India}


\emailAdd{csarkar@scholar.buruniv.ac.in}
\emailAdd{adsarkar@scholar.buruniv.ac.in}
\emailAdd{basundharag@iisc.ac.in}
\emailAdd{bghosh@phys.buruniv.ac.in}
\abstract{With the basic $\alpha$-attractor potentials, we investigate an inflationary regime in the high-$k$ limit, where the cosmological perturbation breaks down due to large enhancement in the scalar power spectrum and generation of  large negative values of the  Bardeen potential. We analyze that, this deep sub-horizon regime creates a situation, which is congenial to the formation of the primordial black holes (PBHs). We work in the spatially flat gauge with $\delta\phi\neq$ 0 and thus explicitly show the roles of perturbations in the inflaton field as well as in the background gravitational field in the mentioned enhancements and thereby in the PBH formation. We calculate the values of $\sigma(M)$, $\beta(M)$ and  $f_\mathrm{PBH}(M)$ around the peaks in the density contrast profile and thus estimate the fraction of PBH in the dark matter of the present universe, corresponding to certain mass scales. We observe the formation of PBHs in the $k$ range $0.43\times 10^{13}$ Mpc$^{-1}$ to $9.8\times 10^{13}$ Mpc$^{-1}$ with masses $1.35\times 10^{-13}M_\odot$ to $2.60\times 10^{-16}M_\odot$, evaporation times $7.74\times 10^{33}$ sec to $5.53\times 10^{25}$ sec, Hawking temperatures $3.72\times 10^{-8}$ GeV to $1.93\times 10^{-5}$ GeV and $f_\mathrm{PBH}(M)$ $\sim  6.12\times 10^{-6}$ to $3.63\times 10^{-1}$. The calculated mass range lies in the regions of forecasts by LISA, WD, NS, DECIGO/AI, FL, SIGWs and the $f_\mathrm{PBH} (M)$ results overlap with those of  DECIGO/AI, FL, SIGWs. }
\maketitle
\flushbottom
\section{Introduction}
\label{sec:intro}
 The concept of primordial black holes (PBHs), associated with comparatively large density fluctuations at the early stages of the evolution of the universe, was introduced by Zel'dovich and Novikov \cite{Zeldovich:1967lct} and later theorized in details by S. W. Hawking and B. J. Carr in \cite{Hawking:1971ei, Carr:1974nx, Carr:1975qj}. Subsequently, the hydrodynamics of PBH formation \cite{1978SvA....22..129N}, accretion of matter around PBHs \cite{1979A&A....80..104N} and PBH formations in the contexts of Grand Unified Theories \cite{Khlopov:1980mg} and in a double inflation in supergravity \cite{Kawasaki:1997ju} were studied. It is conjectured that these PBHs have either evaporated by Hawking radiation or have evolved into supermassive  black holes \cite{Kawasaki:2012wr} or remain as dark matter (DM) in the present universe  \cite{Kawasaki:2012wr,Carr:2020xqk, Villanueva-Domingo:2021spv,Conzinu:2020cke,MoradinezhadDizgah:2019wjf,Clesse:2016vqa,Bird:2016dcv,Kovetz:2017rvv,Inomata:2017okj,Bringmann:2018mxj,Raidal:2018bbj,Green:2020jor,Poulin:2017bwe,Wong:2020yig,Escriva:2021pmf,Calabrese:2021zfq}. The relations between PBHs and primordial gravitational waves have been studied in \cite{Clesse:2016vqa,Choudhury:2013woa, Kovetz:2016kpi,Nakama:2016gzw,Sasaki:2016jop,Kovetz:2017rvv,DeLuca:2020qqa,Domenech:2021wkk,Ozsoy:2020kat,Kimura:2021sqz,Niikura:2017zjd,Katz:2018zrn}. Formation of PBH during a first order phase transition in the inflationary period has also been studied \cite{Khlopov:2008qy,DeLuca:2021mlh}. Studies of the influence of PBHs on the CMB $\mu$ and $y$  distortions  \cite{Deng:2020pxo,Tashiro:2008sf} and $\mu T$ \cite{Ozsoy:2021qrg} cross-correlations have been carried out. Detections of signals from gravitational wave (GW) background, connected with PBH formations, in present and future experiments, have been discussed in Refs. \cite{Garcia-Bellido:2016dkw, Braglia:2020taf,Kawamura:2020pcg}.
\par
In some of the studies of the inflationary scenarios, PBHs have been identified as massive compact halo object (MACHO) with mass $\sim 0.5 M_ {\odot}$ in the work by J. Yokoyama \cite{Yokoyama:1999xi}, who has also examined the formation of PBHs in the framework of a chaotic new inflation \cite{Yokoyama:1998pt}. Josan and Green \cite{Josan:2010cj},  studied constraints on the models of inflation through the formation of PBHs, using a modified flow analysis. Harada, Yoo and Kohri \cite{Harada:2013epa} examined the threshold of PBH formation, both analytically and numerically. R. Arya \cite{Arya:2019wck} has considered the PBH formation as a result of  enhancement of power spectrum during the thermal fluctuations in a warm inflation.
 Formation of PBHs in density perturbations was studied in two-field Hybrid inflationary models \cite{Garcia-Bellido:1996mdl,Chongchitnan:2006wx}, Starobinsky model with a non-minimally coupled scalar field \cite{Pi:2017gih} and the same model including dilaton \cite{Gundhi:2020kzm}, multi-field inflation models \cite{Palma:2020ejf}, isocurvature fluctuation and chaotic inflation models \cite{Yokoyama:1999xi}, inflection-point models \cite{Choudhury:2013woa,Ballesteros:2017fsr,Bhaumik:2019tvl}, quantum diffusion model \cite{Biagetti:2018pjj}, model with smoothed density contrast in the super-horizon limit \cite{Young:2019osy} and with the collapse of large amplitude  metric perturbation \cite{Musco:2018rwt} and large-density perturbation \cite{Young:2020xmk}  upon horizon re-entry. PBH abundance in the framework of non-perturbative stochastic inflation has been studied by F. K\"uhnel and K. Freese \cite{Kuhnel:2019xes}. Relation between the constraints of primordial black hole abundance and those of the primordial curvature power spectrum has also been studied \cite{Kalaja:2019uju,Dalianis:2018ymb}. Recently, PBHs solutions have been obtained   in the framework of non-linear cosmological perturbations and non-linear effects arising at horizon crossing \cite{Musco:2020jjb}.\par
PBH production has recently been studied, in the framework of $\alpha$-attractor polynomial super-potentials and modulated chaotic inflaton potentials models \cite{Dalianis:2018frf}. Mahbub \cite{Mahbub:2019uhl} utilized  the superconformal inflationary $\alpha$-attractor potentials with a high level of fine tuning to produce an ultra-slow-roll region, where the enhancement for curvature power spectra giving rise to massive PBHs was found at $k \sim  10^{13}$ Mpc$^{-1}$. In a subsequent work \cite{Mahbub:2021qeo}, this author re-examined  the earlier work using the optimised peak theory. The ultra-slow-roll process along with a non-Gaussian Cauchy probability distribution has been applied in Ref. \cite{Biagetti:2021eep} to obtain large PBHs masses. The constant-rate ultra-slow-roll-like inflation \cite{2021hllNg:} has also been applied to obtain the enhancement in the power spectra, triggered by entropy production, resulting in PBH formation. Ref. \cite {Teimoori:2021pte} has simulated the onset of PBHs formation by adding a term to the non-canonical $\alpha$-attractor potential, which enhances the curvature perturbations at some critical values of the field. The enhancement of the power spectrum by a limited period of strongly non-geodesic motion of the inflationary trajectory and consequent PBH production has been studied by J. Fumagalli \textit{et al.} \cite{Fumagalli:2020adf}.
\par
In the present paper, our motivation is to explore the possibility of formation of PBHs as a result of breakdown of perturbation, both in the inflton field as well as in the background gravitational field, in the high $k$ sub-horizon regime soon after the end of inflation. As described above, the PBH formation, so far, has been studied in terms of the curvature perturbation $\mathcal{R}$ and the curvature power spectrum $P_{\mathcal{R}} (k)$, which are usually obtained from the Mukhanov-Sasaki equation \cite{1988ZhETF..94....1M,MUKHANOV1992203,mukhanov_2005,10.1143/PTP.76.1036} in the co-moving gauge, characterised by a zero inflaton perturbation ($\delta\phi = 0$). This way of analysis ignores the role of the inflaton perturbation and its possible breakdown in the event of PBH formation. In the present work, we shall examine an alternative route. We shall use the spatially flat gauge, thereby including the role of $\delta\phi$ and the Bardeen potential, $\Phi_{B}$, in the mechanism of PBH formation in $k$ space. In this respect, we shall follow the formalism developed in our previous work \cite{Sarkar:2021ird}, comprising a set of coupled non-linear perturbative evolution equations which could explain the Planck-2018 data \cite{Planck:2018jri} in low $k$ limit. We shall show, here, that the same equations can yield PBH-like solutions in the high $k$ regime with the chaotic $T$ and $E$ model potentials. We find that, so far as the high $k$ limit is concerned, the perturbations in the $k$ space begin to break down at some point in the deep sub-horizon regime. This happens together in the inflaton field as well as in the background gravitational field. In fact, we shall highlight here the important role played by the $\Phi_B$ in building up the density contrasts and the associated PBH formations, which, to our knowledge, has not been done so far in the literature. We believe that this work will open up an avenue for the dynamical origin of PBH formation in the deep sub-horizon $k$-space.\par 
The paper is organised as follows. In Section \ref{subsec:1} we briefly describe the basic formalism of the linear perturbation theory,  leading to the setting up of the three coupled non-linear differential equations which play the central role of our study of the PBH formations. In Section \ref{subsec:alpha_attractor} we write about the basic $\alpha$-attractor $T$ and $E$ model potentials which have been used in the present study. The expression of the transfer function is written and elaborated in Section \ref{subsec:trans_function}.  In Section \ref{subsec:breakdown of perturbation} we explain the mechanism of breakdown of perturbation in the high $k$ limit, which eventually  yields the enhancements in the density contrast in radiation dominated era.  In Section \ref{subsec: dark matter} we do some statistical analysis regrading the connection between the PBH and the DM, around the points of enhancements in the density contrast in the  $k$-space. Results and discussion are presented in Section \ref{sec:results}. Finally, in Section \ref{sec:conclusions} we make some concluding remarks.
\section{Formalism}
\label{sec:Formalism}
\subsection{Linear perturbations in the metric and the inflaton field}
\label{subsec:1}
The Einstein-Hilbert action with minimal coupling between quantised inflaton field,
\begin{equation}
    \phi(t,\vec{X})=\int\frac{d^3 k}{(2\pi)^3}[\phi(k,t)\hat{a}(\vec{k})e^{i\vec{k}.\vec{x}}+\phi^* (k,t)\hat{a}^\dagger (\vec{k})e^{-i\vec{k}.\vec{x}}],
    \label{equ:firstequ}
\end{equation}
and the background linearly-perturbed metric, in spatially flat gauge, with no anisotropic stress,
\begin{equation}
    ds^2 =-(1+2\Phi)dt^2 +2a(t)\partial_i B dx^i dt +a^2 (t)\delta_{ij}dx^i dx^j,
    \label{equ:secequ}
\end{equation}
is
\begin{equation}
    S=\int d^4 x \sqrt{-g}\left(\frac{1}{2}R -\frac{1}{2}g^{\alpha\beta}\partial_{\alpha}\phi\partial_{\beta}\phi -V(\phi)\right).
\end{equation}
The linear perturbation in the inflaton field is written as
\begin{equation}
    \phi(t,\vec{X})=\phi^{(0)}(t) + \delta\phi(t,\vec{X}).
\end{equation}
The perturbation can be translated, through the energy-momentum tensor to that in the density as,
\begin{equation}
    \rho(t,\vec{X})=\rho^{(0)}(t)+\delta\rho(t,\vec{X}),
\end{equation}
where,
\begin{equation}
    \rho^{(0)}(t)=\frac{{\dot{\phi}}^{(0)^2}}{2}+V(\phi^{(0)})
    \label{eq:unperturbed_density}
\end{equation} and
\begin{equation}
    \delta\rho(t,\vec{X})=\frac{dV(\phi^{(0)})}{d\phi^{(0)}}\delta\phi + \dot{\phi}^{(0)}\delta\dot{\phi}-\Phi{\dot{\phi}}^{(0)^2}.
    \label{eq:density_perturbation}
\end{equation}
(Note: The last term in Eq.(\ref{eq:density_perturbation}) was neglected in Ref. \cite{Sarkar:2021ird} as $\Phi{\dot{\phi}}^{(0)^2}$ is small in slow-roll approximation. In the present paper, we retain this term as we expect the metric perturbation $\Phi$ to play a major role in the PBH formation.)
Using the  solutions in \cite{Baumann:2009ds} of the unperturbed Einstein's equations, 
\begin{equation}
    H^2=\frac{\rho^{(0)}}{3},
\label{eq: equation_1}    
\end{equation}
\begin{equation}
\dot{H}+H^2=-\frac{1}{6}(\rho^{(0)}+3p^{(0)})
\label{eq: equation_2}
\end{equation}
and the perturbed Einstein's equations,
\begin{equation}
    3H^2\Phi +\frac{k^2}{a^2} \left(-aHB \right) =-\frac{\delta\rho}{2},
\label{eq:equation_3} \end{equation}
\begin{equation}
    H\Phi=-\frac{1}{2}\delta q,
\label{eq:equation_4}\end{equation}
\begin{equation}
    H\dot{\Phi}+(3H^2+2\dot{H})\Phi=\frac{\delta p}{2},
\label{eq:equation_5} \end{equation}
\begin{equation}
    (\partial_t +3H)\frac{B}{a}+\frac{\Phi}{a^2}=0
\label{eq:equation_6} \end{equation}
and the Bardeen potentials \cite{PhysRevD.22.1882} 
\begin{equation}
    \Phi_B = \Phi-\frac{d}{dt}\left[a^2\left(-\frac{B}{a}\right)\right],
\label{eq:equation_14} 
\end{equation}
\begin{equation}
    \Psi_B = a^2H\left(-\frac{B}{a}\right)
\label{eq:equation_15} 
\end{equation}
we obtain a relation between $\Phi$ and $\Phi_B$ as,
\begin{equation}
    \Phi=\Phi_B+\partial_t \left(\frac{\Phi_B}{H}\right).
\label{eq:equation_20}    
\end{equation}
In Eqs. (\ref{eq:equation_4}) and (\ref{eq:equation_5}) $\delta q$ and $\delta p$ are the magnitudes of the momentum perturbation and pressure perturbation, respectively.
Eqs. (\ref{eq:density_perturbation}) and (\ref{eq:equation_20}) show that the density perturbation $\delta\rho$ contains $\phi^{(0)}$, $\delta\phi$ as well as the Bardeen potential $\Phi_B$.\par 
\par Using the slow roll dynamical horizon crossing condition,  $k=aH$, we go from the space of the cosmic time $t$ to that of the mode momentum $k$ and set up three nonlinear coupled differential equations in the $k$-space and solve for the quantities, $\phi^{(0)}(k)$, $\delta\phi(k)$ and $\Phi_B (k)$. The equations, similar to those in \cite{Sarkar:2021ird}, are,
\begin{equation}
    \delta\phi (k^2 \phi'' + k^2 G_1 \phi'^2 + 4k\phi'+ 6G_1)+ \delta\phi' (-2k^3 G_1 \phi'^2)=0,
    \label{eq:17}
\end{equation}
\begin{multline}
     \delta\phi(1+12G_1 ^2 + 6G_2)+ \delta\phi' (4k+ k^2 G_1 \phi')+k^2\delta\phi''+ \Phi_B (-k\phi'+12G_1 \\+ k^2 G_1 \phi'^2 + k^3 G_1 \phi' \phi'' -12kG_1 ^2 \phi'  +k^3 G_2 \phi'^3)+ \Phi'_B (-2k^2\phi' + 12kG_1 +k^3 G_1 \phi'^2) \\+ \Phi''_B (-k^3\phi')=0
     \label{eq:18}
\end{multline}
and
\begin{multline}
  \Phi''_B \left(-\frac{k^4}{3} \phi'^2\right) + \Phi'_B \left(-\frac{k^3}{3}\phi'^2 + \frac{k^4}{3} G_1 \phi'^3 - k^3 \phi'^2  -\frac{2k^4}{3}\phi'\phi'' + \frac{2k}{3}+ 2k^3\phi'^2 \right)\\ + \Phi_B\left (-\frac{2k^2}{3}\phi'^2 - \frac{2k^3}{3}\phi'\phi'' + k^3 G_1 \phi'^3+\frac{k^4}{3}G_2 \phi'^4 + k^4 G_1 \phi'^2 \phi'' -2 + 2k^2 \phi'^2 -2k^3 G_1\phi'^3\right)\\+\delta\phi''\left(\frac{k^3}{3}\phi'\right)+\delta\phi'\left(2kG_1+\frac{2k^2}{3}\phi' + \frac{k^3}{3}\phi''\right)+\delta\phi \left(2kG_2 \phi' + 2k^2 \phi'' + 2k^2G_1 \phi'^2 +2k\phi'\right)\\=0, 
  \label{eq:19}
\end{multline}

where, $G_n = \frac{\partial^n}{\partial\phi^{(0)^n}}\ln\sqrt{V(\phi^{(0)})}$, $n=1, 2$.
\par In the slow-roll approximation, the density contrast in $k$-space, which is useful for the study of the PBH formation, is given by,
\begin{eqnarray}
\frac{\delta\rho (k)}{\rho^{(0)}(k)}=\nonumber\frac{\delta V (k) +V^{(0)}(k)\left( \frac{k^2}{3}\phi'\delta\phi' + \Phi_B \left(-\frac{k^2}{3}\phi'^2 + \frac{k^3}{3}G_1 \phi'^3\right) + \Phi'_B \left(-\frac{k^3}{3}\phi'^2 \right)\right)}{V^{(0)}(k)}\\=2G_1 \delta\phi + \frac{k^2}{3}\phi'\delta\phi' + \Phi_B \left(-\frac{k^2}{3}\phi'^2 + \frac{k^3}{3}G_1 \phi'^3\right) + \Phi'_B \left(-\frac{k^3}{3}\phi'^2 \right)
\label{eq:density_contrast}
\end{eqnarray}
where,
\begin{equation}
 \delta V(k) \equiv 2G_1 V^{(0)}(k)\delta\phi(k)  
 \label{equ:deltavk}
\end{equation}
and  $V^{(0)}(k) \equiv  V(\phi^{(0)}(k))$.\\ In first line of Eq. (\ref{eq:density_contrast}), we have used the Fourier-space  version of the slow-roll approximation and, therefore, following Eq. (\ref{eq:unperturbed_density}), we have written, $\rho^{(0)}(k)\approx V^{(0)}(k)$. We will show that, both $\delta\phi(k)$ and $\Phi_B (k)$ in Eq. (\ref{eq:density_contrast}) will play a significant role in making $\frac{\delta\rho(k)}{\rho^{(0)}(k)}>\Delta_c (\approx 0.41)$, \textit{i.e.} in the formation of PBHs in the early universe, $\Delta_c$ being the density contrast (see Section \ref{sec:results} for details).
\subsection{The \texorpdfstring{$\alpha$}{a}-attractor \texorpdfstring{$E$}{E}-model and \texorpdfstring{$T$}{T}-model potentials}
\label{subsec:alpha_attractor}
 We use here, the basic $\alpha$-attractor potentials \cite{Carrasco:2015rva,Kallosh:2013yoa}:
\begin{itemize}
 \item[(I)] the $T$-model potential
 \begin{equation}
      V(\phi)=V_{0} \tanh^n\frac{\phi}{\sqrt{6\alpha}},
      \label{eq:T model}
 \end{equation}
 \item[(II)] the $E$-model potential
 \begin{equation}
     V(\phi)=V_{0}(1-e^{-\sqrt{\frac{2}{3\alpha}}\phi})^n
     \label{eq: E model}
 \end{equation}
\end{itemize}
 where $\alpha$ is the inverse curvature of $SU(1,1)/U(1)$ K\"{a}hler manifold \cite{Kallosh:2013yoa}.
 \subsection{Transfer function}
 \label{subsec:trans_function}
 We have related the quantities at horizon crossing  during inflation to those at horizon re-entry in the radiation-dominated era using the transfer function given in Ref. \cite{Musco:2020jjb}:
 \begin{equation}
     T(k,\eta)=3\frac{\sin{(k\eta}/\sqrt{3})-(k\eta/\sqrt{3})\cos{(k\eta/\sqrt{3}})}{(k\eta/\sqrt{3})^3},
 \label{eq:transfer function}   
 \end{equation}
  where $\eta$ is the conformal time.
 \begin{figure}[H]
	\centering
	\includegraphics[width=0.8\linewidth]{"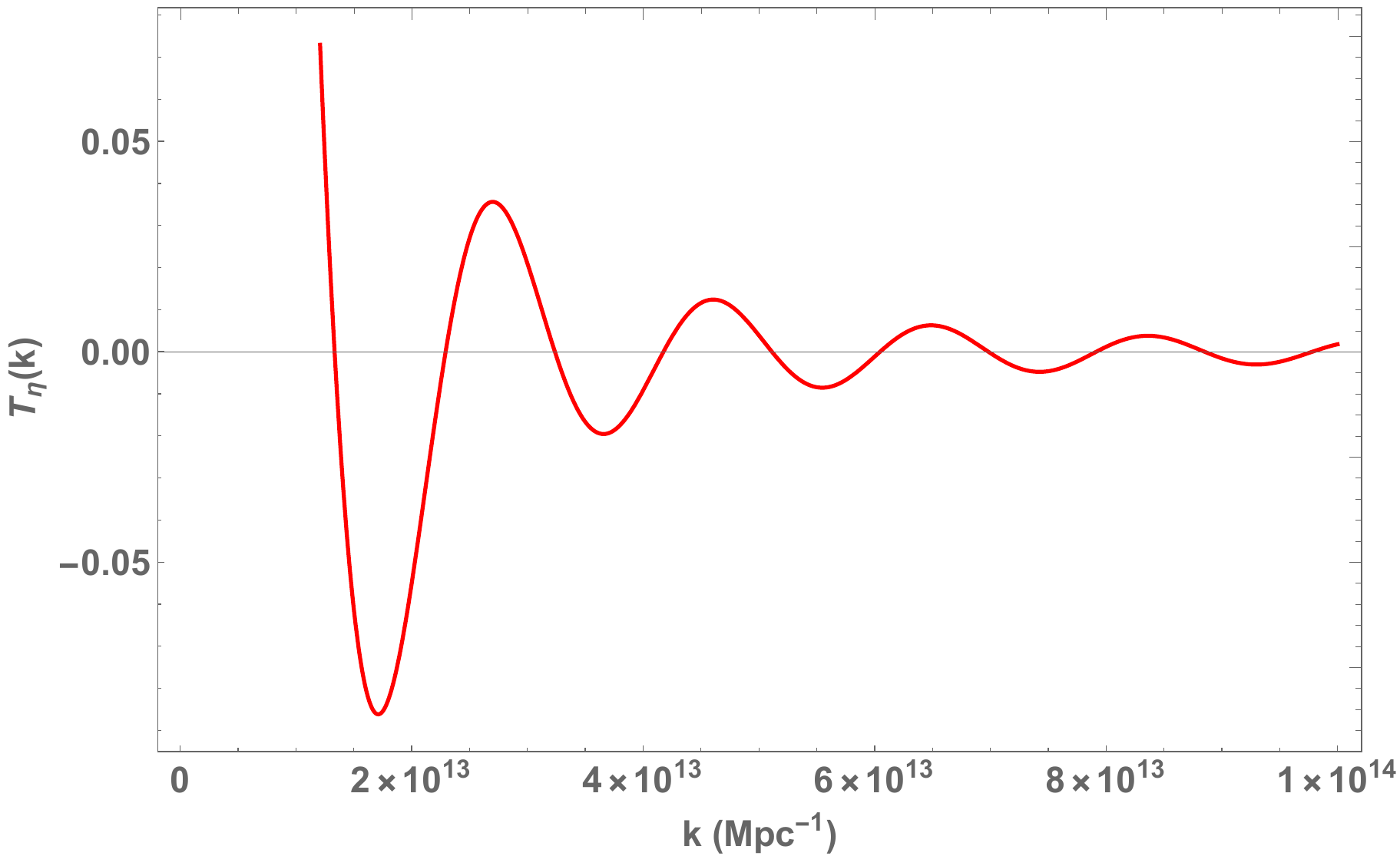"}
	\includegraphics[width=0.8\linewidth]{"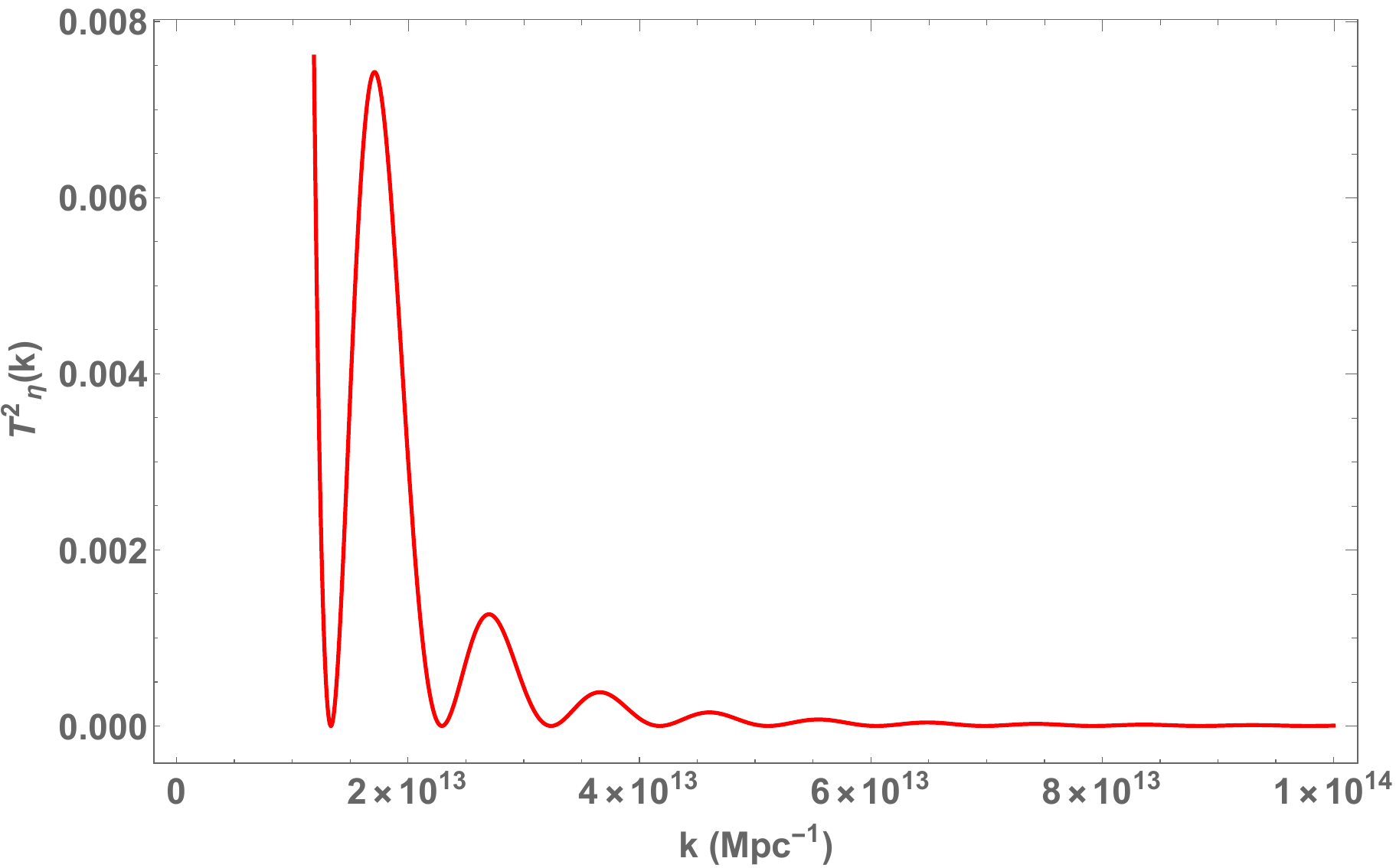"}
	\caption{Transfer function and its square for radiation dominated period at a very small conformal time $\eta \approx 10^{-12}$. It has a very small oscillatory behaviour around $k\sim 6\times10^{13}$ Mpc$^{-1}$, where $k\eta> 1$, although it dies down in the high $k$ limit, $k\rightarrow\infty$. The momentum scale reflects the range of sub-horizon momenta which will be studied in this paper. We will see in Figures \ref{fig:Density_1}, \ref{fig:Bardeen_1} and Table \ref{tab:table_name_fbh} under Section \ref{sec:results}, that this oscillatory behaviour of the transfer function and its square in this momentum range will provide important statistical properties of the PBHs.}
	\label{fig:Transfer_function}
\end{figure}
The transfer function in radiation-dominated era, mainly, creates the pressure gradient  and smooths out the sub-horizon modes \cite{Musco:2020jjb,Baumann:2009ds}. In the present work, we multiply the square of the transfer function with density contrast and go to the radiation-dominated era in which the PBHs are formed. In the $k$ space, the transfer  function gives the required enhancements to the density contrast above the critical value which is necessary for the formation of the PBH. The transfer function in Eq.(\ref{eq:transfer function}) suggests  that the PBHs are formed at a conformal time $\eta$ $\approx$ $10^{-12}$ in the $k$ range $0.43\times10^{13}$ to $9.8\times10^{13}$ Mpc$^{-1}$ which is compatible with the forecasts of Laser Interferometer Space Antenna (LISA), Femtolensing (FL), White Dwarf (WD), Neutron Stars (NS), Scalar Induced Gravitational Waves (SIGWs) Deci-hertz Interferometer Gravitational wave Observatory (DECIGO) experimental projects.    
\subsection{Breakdown of perturbation}
\label{subsec:breakdown of perturbation}
We consider, here, breakdown of cosmological perturbation in the $k$-space in the high $k$ limit \textit{i.e.} in the sub-horizon regime in the spatially flat gauge . Earlier, this breakdown was studied as a consequence of the  effect of higher dimensional operators on the coupled inflaton-metric system \cite{Armendariz-Picon:2008pwa} with $\delta\phi$ $\neq$ $0$. In the present work, the breakdown of perturbation and the associated large contributions to the scalar power spectrum and the  background gravitational field (see Figures \ref{fig:Power_spectrum} and \ref{fig:Bardeen_1}) can be linked with the terms involving higher powers of $k$ ($k^3$ and $k^4$) which are the dominant terms in Eqs.(\ref{eq:17}), (\ref{eq:18}) and (\ref{eq:19}) in the high $k$ limit. Higher powers of $k$ arise in these equations because of the mixing of the modes of $\phi^{(0)}(k)$, $\delta\phi(k)$ and $\Phi_B (k)$. The modes of $\Phi_B(k)$ are classical Fourier modes, whereas that of $\phi^{(0)}(k)$ and $\delta\phi(k)$ are of quantum origin. This mixing of the modes signifies the interaction of the quantized inflaton field with the classical background gravitational field.  \par
In earlier studies, it was shown that the PBHs are formed with substantial amount in the over-dense regions when, after horizon re-entry, large classical density perturbations above the threshold $\Delta=\frac{\delta\rho}{\rho}>\Delta_c$ collapse against outward radiation pressure \cite{PhysRevD.50.7173,Garcia-Bellido:1996mdl,Ivanov:1997ia,PhysRevD.96.043504}, with mass of the order of the horizon mass \cite{Teimoori:2021pte},
\begin{equation}
    M_\mathrm{PBH}=\gamma M_\mathrm{hor}, \quad\mathrm{with}\quad\gamma\approx 0.2.
\end{equation} This mass can be related to the inflationary co-moving mode momentum $k$ as \cite{Teimoori:2021pte},
\begin{equation}
M_\mathrm{PBH}(k)=3.68\left(\frac{\gamma}{0.2}\right)\left(\frac{g_*}{10.75}\right)^{-1/6}\left(\frac{k}{10^6 \mathrm{Mpc}^{-1}}\right)^{-2}\mathrm{M_\odot} \quad \mathrm{with}\quad g_*=107.5.
\label{eq:massPBH}
\end{equation}Here, $M_\mathrm{PBH}$ is a monochromatic PBH mass distribution function as we are not considering physical processes like accretion, merger etc.(see Ref. \cite{Bellomo:2017zsr} for details).\par 
The  formation of Schwarzschild-type PBH is usually constrained by the rate $\beta (M)$ and the abundance $f_\mathrm{PBH} (M)$ with respect to a Gaussian probability distribution $P(\Delta;M)$ of density fluctuations, resulting from curvature perturbation $\mathcal{R}$ in the Press-Schechter formalism \cite{Press:1973iz}, given by,
\begin{equation}
    P(\Delta;M)=\frac{1}{\sqrt{2\pi\sigma^2}}e^{-\frac{\Delta^2}{2\sigma^2}}
    \label{eq:probability},
\end{equation} where $\sigma^2$ is referred to as the coarse-grained variance of density contrast for the PBH of mass $M$ \cite{Young_2014}. The required amount of density contrast is  obtained by introducing a geometric modification of inflaton potential, called ultra-slow roll (USR), which helps in enhancement of the primordial curvature power spectrum $P_{\mathcal{R}}(k)$ in co-moving gauge. However, as discussed in Section \ref{sec:intro}, this does not  incorporate the role of inflaton perturbation $\delta\phi$ and the Bardeen potential $\Phi_B$ in the PBH formation. On the contrary,  a formalism with spatially flat gauge is able to take care of these important ingredients of PBH formation (see Eq.(\ref{eq:density_contrast})). In this present paper, we work in this formalism (see \cite{Sarkar:2021ird} for details). Our framework yields the needed amount of density fluctuations from enhancements in the  first order perturbation $\delta V (k)$ in the potential in $k$ space, as given by Eq.(\ref{equ:deltavk}) and in the Bardeen potential $\Phi_B (k)$ in the negative direction. In the inflationary regime, $k$  $\sim 10^{-3}$ Mpc$^{-1}$ for which $\delta\phi$ is very small and, hence, the enhancement in $\delta V$ is negligible (see \cite{Sarkar:2021ird} for the corresponding plots). But, in the $k$ space of PBH formation $k\sim 10^{13}$ Mpc$^{-1}$, where $\delta\phi\sim 10^{8}$ (see Figure \ref{fig:perturbed-inflaton-high-k}), by which $\delta V$ is magnified by the order of $10^6$ (see Figure \ref{fig:perturbed-potential-high-k}).  This amplification is dictated  by the leading terms with higher powers of $k$ in Eqs.(\ref{eq:17}), (\ref{eq:18}) and (\ref{eq:19}). From Figure \ref{fig:Bardeen_1}, we can understand that the negative enhancement of metric perturbation \textit{i.e.} the Bardeen potential $\Phi_B (k)$, which acts like a classical gravitational potential in the radiation era, creates a favourable situation of PBH formation in the $k$ space, soon after  horizon re-entry($\eta$ $\sim$ $10^{-12}$). Amplifications of the inflaton potential and the Bardeen potential in the momentum space  can therefore be considered as the key factors of PBH formation.
\subsection{PBH and dark matter}
\label{subsec: dark matter}
Unlike the ordinary stellar black holes (BHs), formed from the collapse of the core of a dying star \cite{PhysRev.55.364,PhysRev.55.374} of mass $M_\mathrm{star}>3M_\odot$ \cite{Chandrasekhar:1964zz},  PBHs could be produced with arbitrary masses \cite{Villanueva-Domingo:2021spv}. PBHs with mass $\lesssim 3M_\odot$ signify the ordinary primordial black holes \cite{Green:2020jor}, but the ones with mass window $10^2 M_\odot-10^5 M_\odot$ carry the signature of PBH mergers \cite{LIGOScientific:2020iuh}. Also, PBH could give rise to the seeds for supermassive black holes (SMBHs), present in the active galactic nuclei (AGNs) with mass range $10^5 M_\odot-10^{10} M_\odot$ and redshift $z>6$ \cite{Carr:2018rid}.\par
Now, PBHs lighter than $M_c \simeq (\frac{3\hbar c^4 \alpha_0 t_0}{G^2})^{1/3}\sim 10^{15} (\frac{\alpha_0}{4\times 10^{-4}})^{1/3}(\frac{t_0}{13.8  \mathrm{Gyr}})^{1/3}\mathrm{g}$ \cite{PhysRevD.13.198} do not exist in the present universe as they have  evaporated by  the Hawking radiation. The PBHs heavier than $M_{c}$ are still present and they are of great interest in cosmology. It is possible that these PBHs comprise all or a fraction of DM \cite{Flores:2021jas}. Given the fact that the DM candidates in particle physics (e.g, WIMPs)  have not been experimentally found yet, this possibility remains quite high. Thus, whether the PBH could be an essential DM candidate or not, is a very pertinent question \cite{2014MPLA...2940005B,Chapline:1975ojl,Carr:1975qj,Clesse:2017bsw,Carr:2016drx, Carr:1976zz,Meszaros:1975ef,Khlopov:2008qy,Frampton:2010sw,Belotsky:2014kca,Carr:2009jm,Wang:2020uvi,Montero-Camacho:2019jte,Laha:2019ssq,Dasgupta:2019cae,Laha:2020ivk,Graham:2015apa,PhysRevD.94.083504,PhysRevD.81.104019,Carr:1974nx}. In fact, it has been found that PBHs have some of the characteristics of a cold dark matter (CDM) \cite{PhysRevD.94.044029,Green:2020jor}. Other evidences of DM to be PBHs of different masses have been found in experiments of BH mergers, gravitational femto-lensing (FL) etc. \cite{Clesse:2016vqa,Villanueva-Domingo:2021spv}. A detailed discussion on various mass windows of PBH abundance in the context of DM has been given in  \cite{Carr:2016drx}. Currently, it appears that PBHs can not explain all of DM, but a fraction of it \cite{PhysRevD.96.023514}. Recently, after the LIGO's declaration of $30 M_\odot$ BH merger event GW150914, many authors pointed out that the merger rate could be explained in terms of  merger rate of PBHs without violating the fact that PBH abundance is equal to or less than the total DM abundance \cite{LIGOScientific:2016vpg,LIGOScientific:2018mvr,LIGOScientific:2017ycc,LIGOScientific:2017vox,LIGOScientific:2017bnn,LIGOScientific:2016aoc,LIGOScientific:2016vlm,LIGOScientific:2016sjg,LIGOScientific:2016kwr,LIGOScientific:2016dsl,PhysRevLett.116.201301,PhysRevLett.117.061101,Nakamura_1997,Bertone:2016nfn,Carr:2020xqk,Green:2020jor,Sasaki:2018dmp,DeLuca:2020qqa,Villanueva-Domingo:2021spv}. A catalog of recent and ongoing experiments on PBH-DM is available in \cite{bradley_j_kavanagh_2019_3538999}.\par 
For the PBHs to be DM candidates, they must have a very large evaporation time ($ t_\mathrm{eva}$) (at least larger than the present age of the universe) and, hence, very small Hawking temperature ($T_\mathrm{H}$). We will analyse these aspects in Section \ref{sec:results} using the equations \cite{Dalianis:2018ymb} given by,
\begin{equation}
    t_\mathrm{eva}=4\times 10^{11} \left(\frac{M_\mathrm{PBH}}{10^{13}}\right)^3 s
    \label{eq:evaporationtime}
\end{equation} and
\begin{equation}
    T_\mathrm{H}=\frac{10^{13}\mathrm{g}}{M_\mathrm{PBH}}\quad\mathrm{GeV}.
    \label{eq:hawkingtemperture}
\end{equation}\par
In the following, we present the way we calculate $f_\mathrm{PBH} (M)$, i.e., the fraction of PBHs present in the total DM.\par
We take the same Gaussian distribution of density fluctuation as in Eq.(\ref{eq:probability}) and calculate the variance of mass distribution corresponding to each peak from Figure \ref{fig:Density_1} as
\begin{equation}
    \sigma^2 (M)=\langle M^2 \rangle - \langle M \rangle ^2
    \label{eq:sigma},
\end{equation} where,
\begin{equation}
    \langle M \rangle = \frac{1}{k_\mathrm{max} - k_\mathrm{min}} \int_{k_\mathrm{min}}^{k_\mathrm{max}}M(k) dk
\end{equation} and 
\begin{equation}
     \langle M^2 \rangle = \frac{1}{k_\mathrm{max} - k_\mathrm{min}} \int_{k_\mathrm{min}}^{k_\mathrm{max}}M^2(k) dk.
\end{equation}
Here, $k_\mathrm{min}, k_\mathrm{max}$ correspond to the values of $k$ at $\Delta$=$\Delta_{c}$.\par Then, we calculate the production rate of PBH, \textit{i.e.}, $\beta (M)$ between the critical value (which is the minimum value $\Delta_\mathrm{min}$) $\Delta_c = 0.42$ and the peak value  $\Delta_\mathrm{max}$ , taken from Figure \ref{fig:Density_1}, as
\begin{equation}
    \beta (M) = \int_{\Delta_\mathrm{min}}^{\Delta_\mathrm{max}}\frac{d\Delta}{\sqrt{2\pi\sigma^2(M)}}e^{-\frac{\Delta^2}{2\sigma^2 (M)}}=\frac{1}{2}\left(\mathrm{erf}\left(\frac{\Delta_\mathrm{max}}{\sigma\sqrt{2}}\right) - \mathrm{erf}\left(\frac{\Delta_{\mathrm{min}}}{\sigma\sqrt{2}}\right)\right).
    \label{eq:rate}
\end{equation}
The fraction of PBHs in the DM halo corresponding to a specific peak  in the density contrast profile, around the peak mass $M_\mathrm{peak}$ is given by \cite{Teimoori:2021pte}
\begin{equation}
    f_\mathrm{PBH}(M)=\frac{\Omega_\mathrm{PBH}}{\Omega_\mathrm{DM}}=\frac{\beta(M)}{3.94\times 10^{-9}}\left(\frac{\gamma}{0.2}\right)^{\frac{1}{2}}\left(\frac{g_{*}}{10.75}\right)^{-\frac{1}{4}}\left(\frac{0.12}{\Omega_\mathrm{DM}h^{2}}\right)\left(\frac{M_\mathrm{peak}}{M_\odot}\right)^{-\frac{1}{2}} 
    \label{eq:fpbh}
\end{equation}
where $\Omega_\mathrm{DM}h^2=0.12$       \cite{Planck:2018jri}, $\Omega_\mathrm{DM}$ and $\Omega_\mathrm{PBH}$ are the dark matter and PBH density respectively. The results of Eqs.(\ref{eq:sigma}), (\ref{eq:rate}) and (\ref{eq:fpbh}) are given in Table \ref{tab:table_name_fbh}.\par 
We may note, here, the difference between the Press-Schecter formalism \cite{Press:1973iz} and our method. In the former, $\sigma^2(k)$, the coarse-grained variance of density contrast is obtained  from the curvature power spectrum and a window function \cite{Dalianis:2018frf,Teimoori:2021pte}, whereas we obtain $\sigma^2(M)$ corresponding to each peak in the density contrast. Our mass variance is of the same order of PBH mass in unit of gram (with $\mathrm{M}_\odot = 1.989\times 10^{33}\mathrm{g}$) (see Table \ref{tab:table_name_fbh}). This parameter reflects the statistical behaviour of PBH mass distribution in $k$ space.

\section{Results and Discussion}
\label{sec:results}
At the outset, let us give an overall perspective of the evolution of the modes during the inflationary period vis-\`a-vis the PBH formation. As shown and stated in Ref. \cite{Sarkar:2021ird}, the higher $k$ modes undergo less number of e-folds and thus, while the Hubble sphere shrinks, they exit the horizon at later times. As a consequence, the very high $k$ modes remain in the deep sub-horizon region during the major part of the inflationary period. When the Hubble sphere starts expanding after the end of inflation, the high $k$ modes re-enter the horizon first in small positive conformal times. The smaller $k$ modes re-enter the horizon at later conformal times. Our interest, here, is in the high $k$ regions where, it will be shown that, the inflationary density contrast shoots up, at a number of momentum values, signalling the breakdown of the perturbative framework and creating a condition favourable for the formation of PBHs, when the modes re-enter the horizon.\par Interestingly, this condition is achieved in our study quite naturally by solving the $k$-space evolution equations, with the original $\alpha$-attractor $T$-model and $E$-model potentials with the simple set of parameters: $n = 2$, $\alpha=1$ and $V_0 = 1$ (see Eqs.(\ref{eq:T model}) and (\ref{eq: E model})). These values belong to the range of parameters which has been shown to be efficacious in fitting the Planck data in the ($n_s$ - $r$) plane \cite{Sarkar:2021ird, Planck:2018jri}. The main thrust of our study is, thus, to examine the roll of breakdown of perturbation at high momentum in the formation of PBHs without any modification of the basic $\alpha$-attractor potentials in the field space.
\par
Next we will discuss our results from the self-consistent solutions of the Eqs. (\ref{eq:17}) - (\ref{eq:19}) during inflation, which have been obtained by considering $\phi^{(0)}(k=10^{8}) = 6.21$, $\delta\phi(k=10^{8})=\Phi_B (k=10^{8})=10^{-14}$, $\phi'^{(0)}(k=10^{8})=\delta\phi'(k=10^{8})=\Phi'_B(k=10^{8})=0$, as initial conditions. These initial conditions are consistent with $N=60$, which is the number of $\mathrm{e}$-folds at the end of the inflation. See \cite{Sarkar:2021ird} for details.\par
\begin{figure}[H]
	\centering
	\includegraphics[width=0.8\linewidth]{"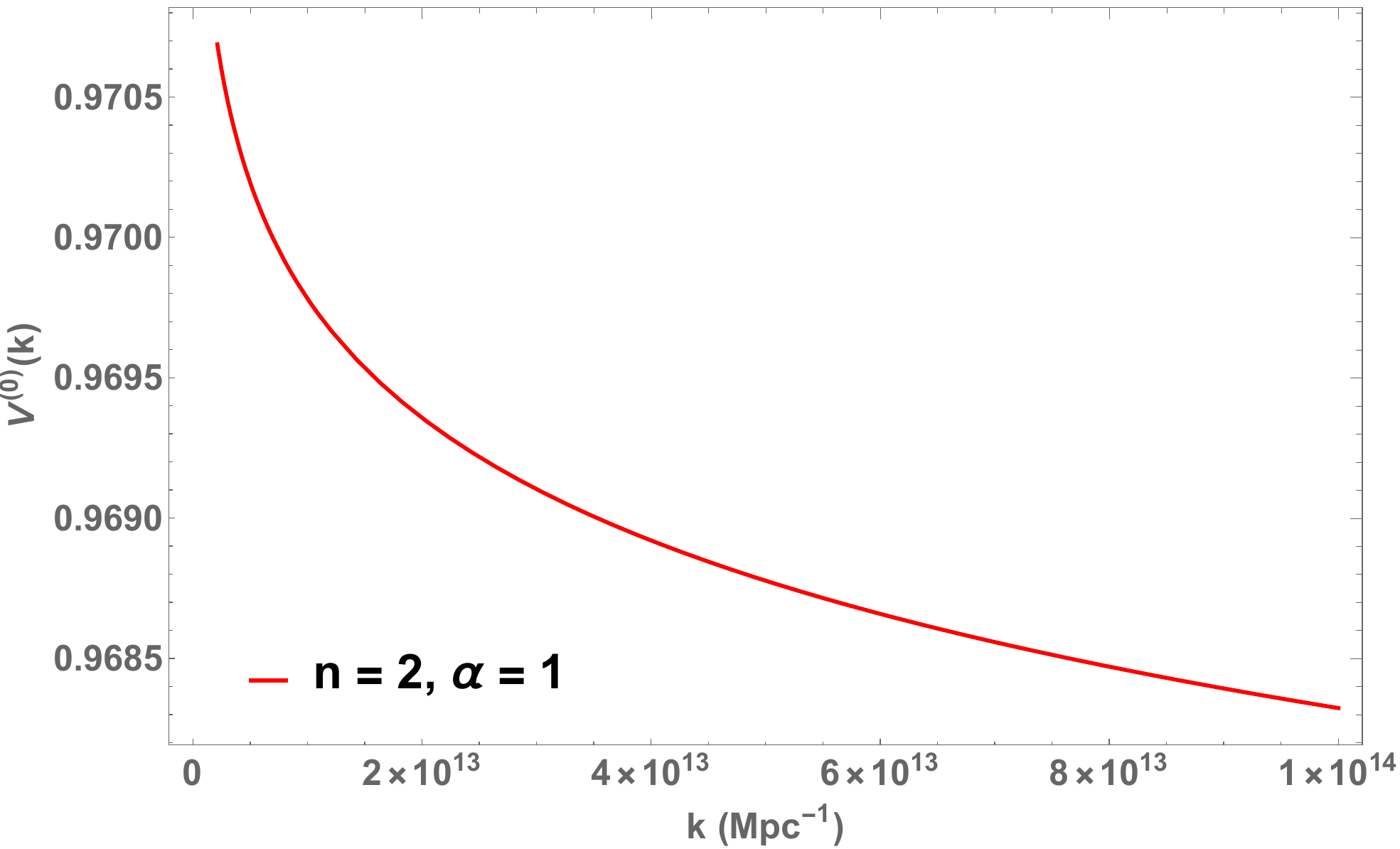"}
	\caption{Unperturbed part $V^{(0)}(k)$ of the chaotic $\alpha$-attractor $T$-model potential at high $k$ limit.}
	\label{fig:potential-high-k} 
\end{figure}

\begin{figure}[H]
	\centering
	\includegraphics[width=0.8\linewidth]{"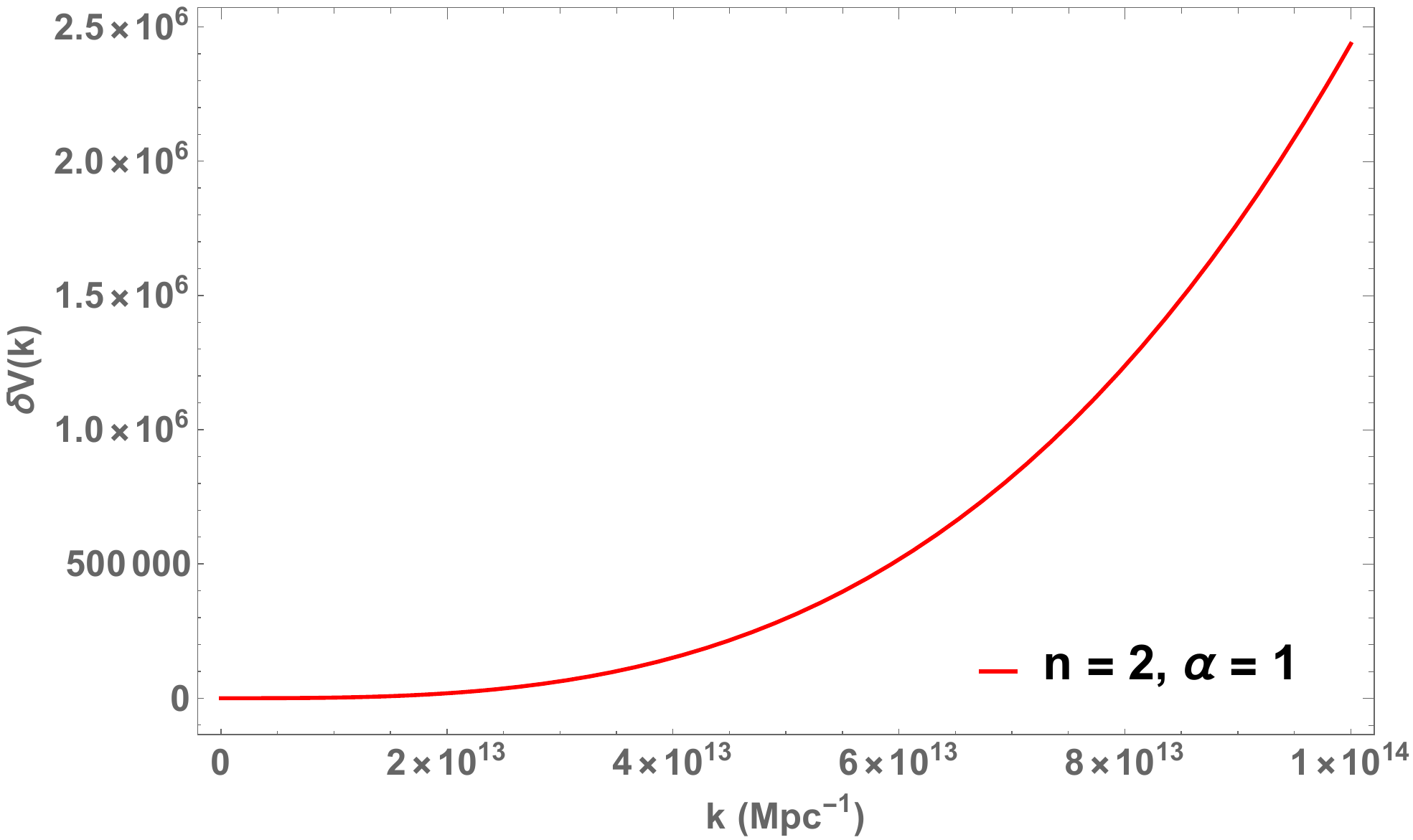"}
	\caption{Perturbation $\delta V (k)$ of chaotic $\alpha$ attractor $T$-model potential in high $k$ limit. The perturbation in the potential, $\delta V(k)\gg V^{(0)}(k)$ (see Figure  \ref{fig:potential-high-k}). It blows up as $k\rightarrow\infty$, signifying the breakdown of the linear perturbation and creating a situation, favourable for the PBH formation.}
	\label{fig:perturbed-potential-high-k}
\end{figure}

\begin{figure}[H]
	\centering
	\includegraphics[width=0.8\linewidth]{"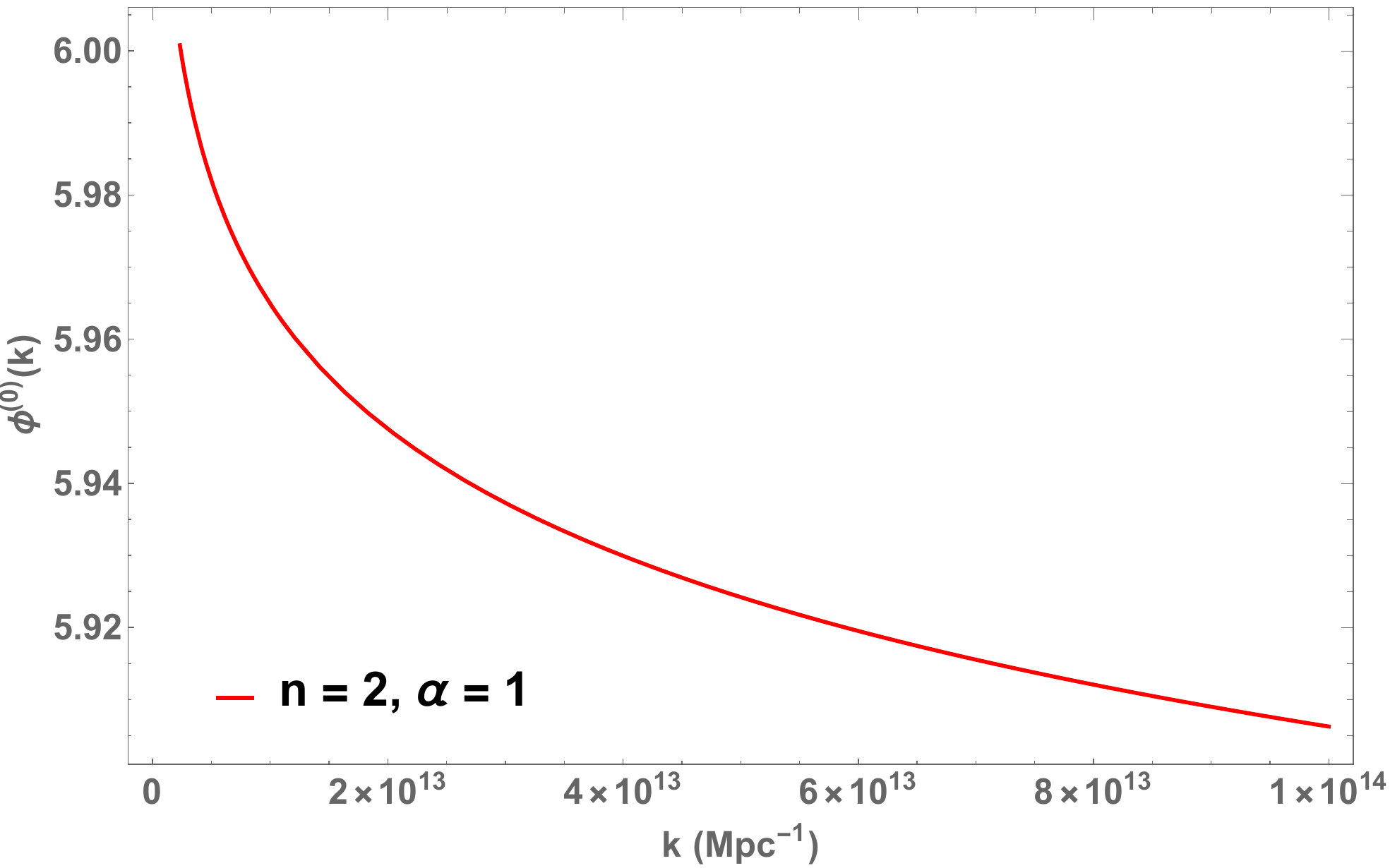"}
	\caption{Unperturbed inflaton field $\phi^{(0)}(k)$ for chaotic $\alpha$ attractor $T$-model potential at high $k$ limit.}
	\label{fig:inflaton-high-k}
\end{figure}

\begin{figure}[H]
	\centering
	\includegraphics[width=0.8\linewidth]{"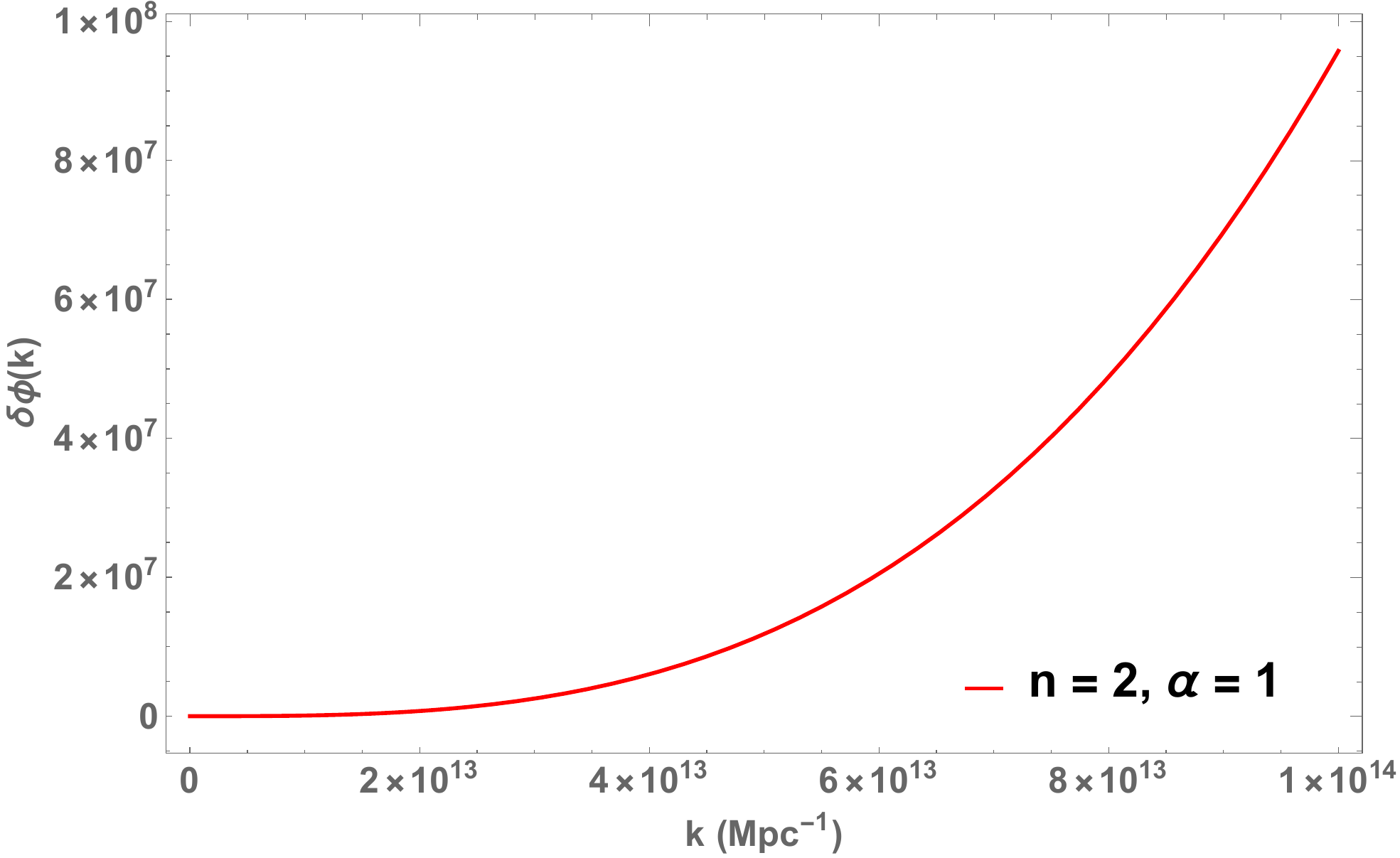"}
	\caption{Perturbation $\delta \phi (k)$ for chaotic $\alpha$ attractor $T$-model potential in high $k$ limit. The perturbation in the inflaton field, $\delta \phi(k)\gg \phi^{(0)}(k)$ (see Figure  \ref{fig:inflaton-high-k}). It blows up as $k\rightarrow\infty$, signifying the breakdown of the linear perturbation and creating a situation, favourable for the PBH formation.}
	\label{fig:perturbed-inflaton-high-k}
\end{figure}

\begin{figure}[H]
	\centering
	\includegraphics[width=0.8\linewidth]{"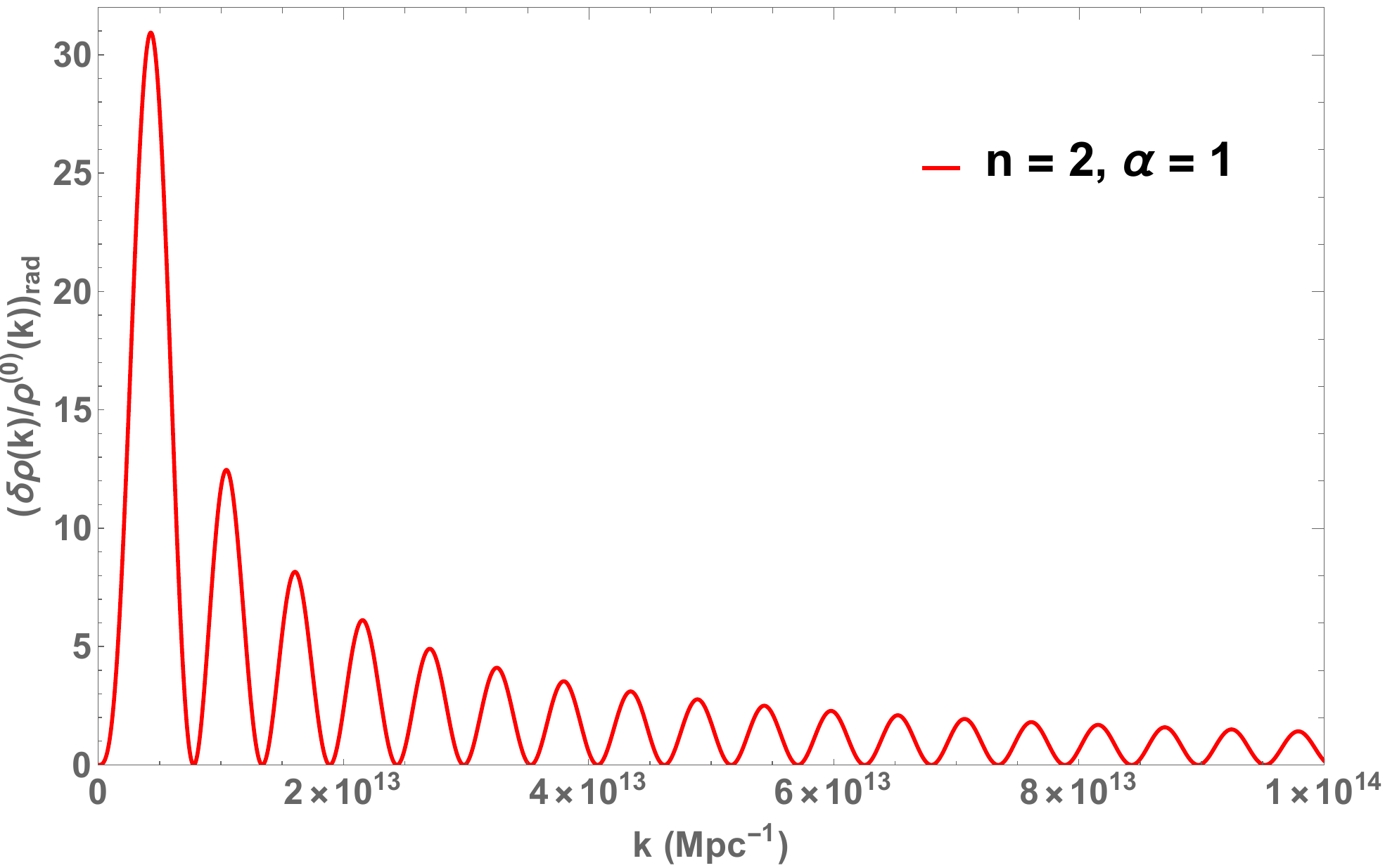"}
	\caption{Density contrast $\left(\frac{\delta \rho(k)}{\rho^{(0)}(k)}\right)_\mathrm{rad}\left[=\left(\frac{\delta \rho(k)}{\rho^{(0)}(k)}\right)_\mathrm{inf}\times T^2(k,\eta=10^{-12}\mathrm{s})\right]$ vs. $k$, in the radiation-dominated era, in a high momentum range ($1.12\times 10^{12}$ Mpc$^{-1}$ to $9.92\times 10^{13}$ Mpc$^{-1}$), where signatures of PBHs have been predicted \cite{Green:2020jor}. Here, $\left(\frac{\delta \rho(k)}{\rho^{(0)}(k)}\right)_\mathrm{inf}$ is the density contrast during inflation. Peaks are observed at $18$ values of $k$. These peaks suggest the presence of PBHs of different masses (see Table \ref{tab:table_name_parameters}).}
	\label{fig:Density_1}
\end{figure}

\begin{figure}[H]
	\centering
	\includegraphics[width=0.8\linewidth]{"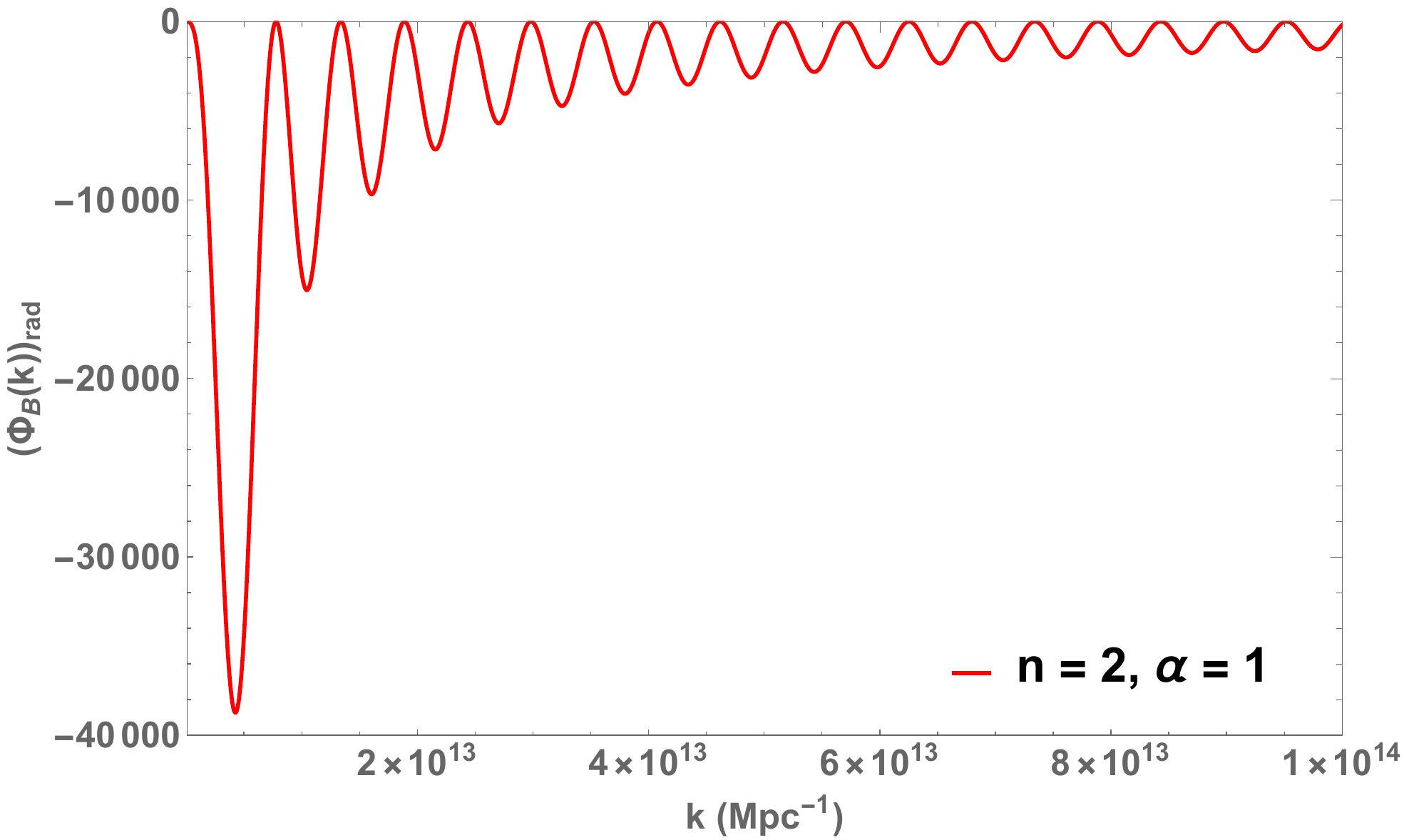"}
	\caption{The Bardeen potential $(\Phi_B(k))_\mathrm{rad}=(\Phi_B(k))_\mathrm{inf}\times T^2(k,\eta=10^{-12})$ vs. $k$ in the radiation-dominated era in the same momentum range as in Figure \ref{fig:Density_1}.  Here, $(\Phi_B(k))_\mathrm{inf}$ is the Bardeen potential during inflation. Large negative values of the potentials are observed at the $k$-values where the enhancements of the density contrast occur (see Figure  \ref{fig:Density_1}). This demonstrates the crucial role played by the Bardeen potential in the formation of the PBHs.}
	\label{fig:Bardeen_1}
\end{figure}

\begin{figure}[H]
	\centering
	\includegraphics[width=0.8\linewidth]{"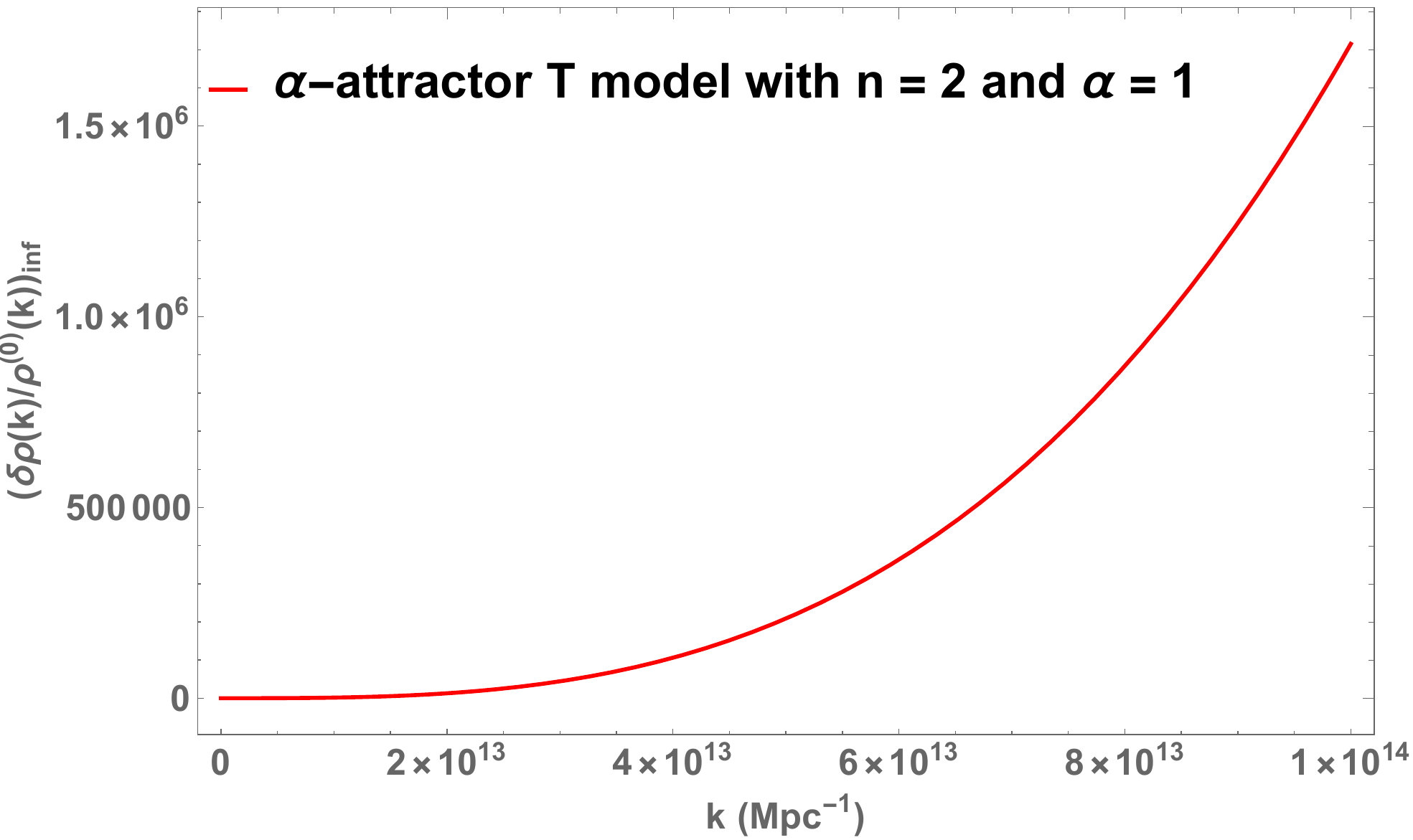"}
	\includegraphics[width=0.8\linewidth]{"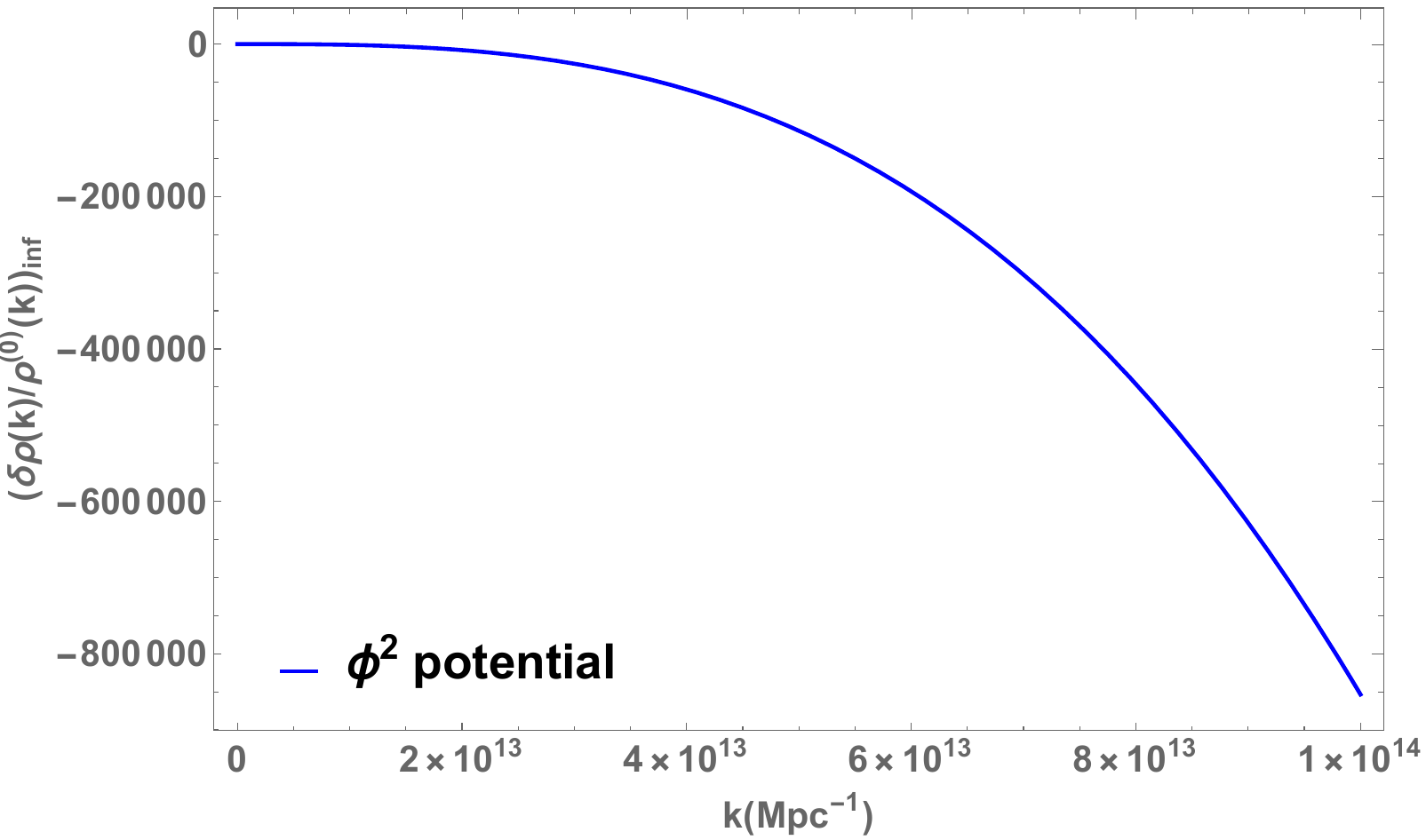"}
	\caption{Comparison of the uses of the $\alpha$-attractor chaotic $T$-model potential (upper figure) and a power law potential, $\phi^2$ (lower figure) in calculating the density contrast $\left(\frac{\delta \rho(k)}{\rho^{(0)}(k)}\right)_\mathrm{inf}$, during inflation, in the high momentum regime. While the density contrast has very large positive values in the former case, it shows large negative values in the latter case. The unphysical nature of the density contrast in high momentum regime, in the case of the power law potential, clearly indicates its unsuitability in the formation of PBHs. Same inference can be drawn during radiation dominated era also.}
	\label{fig:pic2pic4}
\end{figure}

 
In Figure \ref{fig:potential-high-k}, we show the $k$-space behaviour of the unperturbed chaotic $\alpha$-attractor $T$-model potential in the high $k$ limit. The variation of $V^{(0)}(k)\approx 0.9685-0.9705$ is small in the entire range of $k$, which is also observed in low $k$ limit during inflation \cite{Sarkar:2021ird}. However, large amount of change takes place in the perturbed part, $\delta V(k)$, in these momentum range, as shown in the next figure.
In Figure \ref{fig:perturbed-potential-high-k}, we demonstrate the behaviour of the perturbation in the potential, where it is shown that $\delta V(k)$ becomes very large in the high $k$ limit, signaling breakdown of perturbation in this limit. This does not happen in the low $k$ limit, where $\delta V(k)$ remains very small \cite{Sarkar:2021ird}.
In Figure \ref{fig:inflaton-high-k}, we have plotted the unperturbed inflaton field for the chaotic $\alpha$-attractor $T$-model potential in the high $k$ limit and in Figure \ref{fig:perturbed-inflaton-high-k}, the corresponding perturbation, $\delta \phi(k)$, in the same limit. Here also we observe the breakdown of perturbation at high $k$ values. Like Figures \ref{fig:potential-high-k} and \ref{fig:perturbed-potential-high-k}, major change happens in $\delta\phi(k)$ rather than $\phi^{(0)}(k)$ (the variation of $\phi^{(0)}(k)$ is also small in low $k$ limit \cite{Sarkar:2021ird} like $V^{(0)}(k)$). We, thus, realize that the large enhancement in $\delta V$ in momentum space takes place as a result of breakdown of perturbation of the inflaton field, in the same space. 
In Figures \ref{fig:Density_1} and and \ref{fig:Bardeen_1} we plotted the density contrast and the Bardeen potential, respectively, in the radiation-dominated era. These quantities have been obtained by multiplying their corresponding values during inflation with the square of the transfer function Eq. (\ref{eq:transfer function}). Looking at Figure \ref{fig:Density_1}, we observe that the values of the density contrasts at all the peaks are above a threshold value \textit{viz.}, $\Delta_c \cong 0.41$ given in literature \cite{Villanueva-Domingo:2021spv,Musco:2018rwt}. Thus, the peaks satisfy the primary criterion for the PBH formations. It may be noted here that, these peaks correspond to the density contrast (see the upper half of figure in Figure \ref{fig:pic2pic4}, in red color) coming from the self-consistent calculation of $k$-space evolution equations (Eqs. (\ref{eq:17}) -(\ref{eq:19})) and smoothened by the  transfer function, shown in Figure \ref{fig:Transfer_function}.
Figure \ref{fig:Bardeen_1} shows the $k$-space evolution of the Bardeen potential, which  highlights the fact that, the peaks in density contrast correspond to the peaks of the Bardeen potential in the negative direction at the same value of $k= k_\mathrm{peak}$. This result reflects the crucial interplay between the quantum-inflaton-fluctuation-induced density perturbation and the metric perturbation \textit{i.e.}, the Bardeen potential, in the  PBH formation.
Now we can summarise the results of Figures \ref{fig:potential-high-k}-\ref{fig:Bardeen_1} as follows. The enhancements in $\delta\phi (k)$ and therefore in $\delta V(k)$ due to the breakdown of perturbations  in the high $k$ limit, are transferred in the density contrast $\Delta=\frac{\delta\rho (k)}{\rho^{(0)}(k)}$ and the Bardeen potential $\Phi_B (k)$. Consequently, $\Delta$ exceeds the critical \textit{i.e.} threshold value $\Delta_c$ ($\Delta>\Delta_c$) and $\Phi_B (k)$ becomes large negative, simultaneously. Therefore, as described in Section \ref{subsec:breakdown of perturbation}, PBHs are formed from the magnification of the $T$ model $\alpha$ attractor chaotic inflaton potential and the Bardeen potential in the $k$ space.\par 
Figure \ref{fig:pic2pic4} illustrates the fact that the requirement of large positive inflationary density contrasts at high $k$ values for the PBH formations, is satisfied only  by the $\alpha$-attractor $T$-model potential and not by the power law type $\phi^2$ potential, for example. We have also found in \cite{Sarkar:2021ird} that, such type of potential does constitute an experimentally-favourable model for inflation at low $k$ limit within some specified range of parameters. Therefore in $k$ space, the $\alpha$-attractor potential in its pristine form has the capability  of explaining both the inflationary paradigm at the low $k$ limit and the PBH formation at the high $k$ limit. 
Inflationary density contrasts shown in Figure \ref{fig:pic2pic4}, is not of interest because, by definition, density contrast is a classical entity, formed after the horizon re-entry of inflationary perturbations \cite{Baumann:2009ds}.  Therefore the inflationary density contrast must be translated to the radiation era by multiplying with the square of the transfer function, shown in Eq. (\ref{eq:transfer function}) (see Figure \ref{fig:Transfer_function}). We choose the range of the sub-horizon modes $k\sim 10^{13}\mathrm{Mpc}^{-1}$ of PBH formation with conformal time $\eta\sim 10^{-12}$ such that $k\eta>1$. In this $k$ domain, the transfer function and its square show an oscillatory behaviour, as shown in Figure \ref{fig:Transfer_function}. The inflationary density contrast is therefore modulated by the oscillation of square of transfer function as shown in Figure \ref{fig:Density_1}, resulting in a multi-spike distribution pattern of classical density fluctuations over the $k$ modes around different peaks with peak momenta $k_\mathrm{peak}$ and density contrasts $\Delta_\mathrm{max}$ . This $k$ space density profile shows that in the spatially flat gauge PBHs are formed in cluster and for each PBH in this cluster there is a density distribution around a peak value from a minimum value $k_\mathrm{min}$ and maximum value $k_\mathrm{max}$ at $\Delta=\Delta_c$ . We have found $18$ such distributions corresponding to $18$ values of $k_\mathrm{peak}$ around which density contrasts exceed the critical value between $\Delta_\mathrm{min}$ and $\Delta_\mathrm{max}$. Therefore, the transfer function which connects the pre- and post-inflationary regimes at a given conformal time, plays an important role in distributing the density fluctuations among various modes in $k$ space.  
\par
\begin{figure}[H]
	\centering
	\includegraphics[width=0.8\linewidth]{"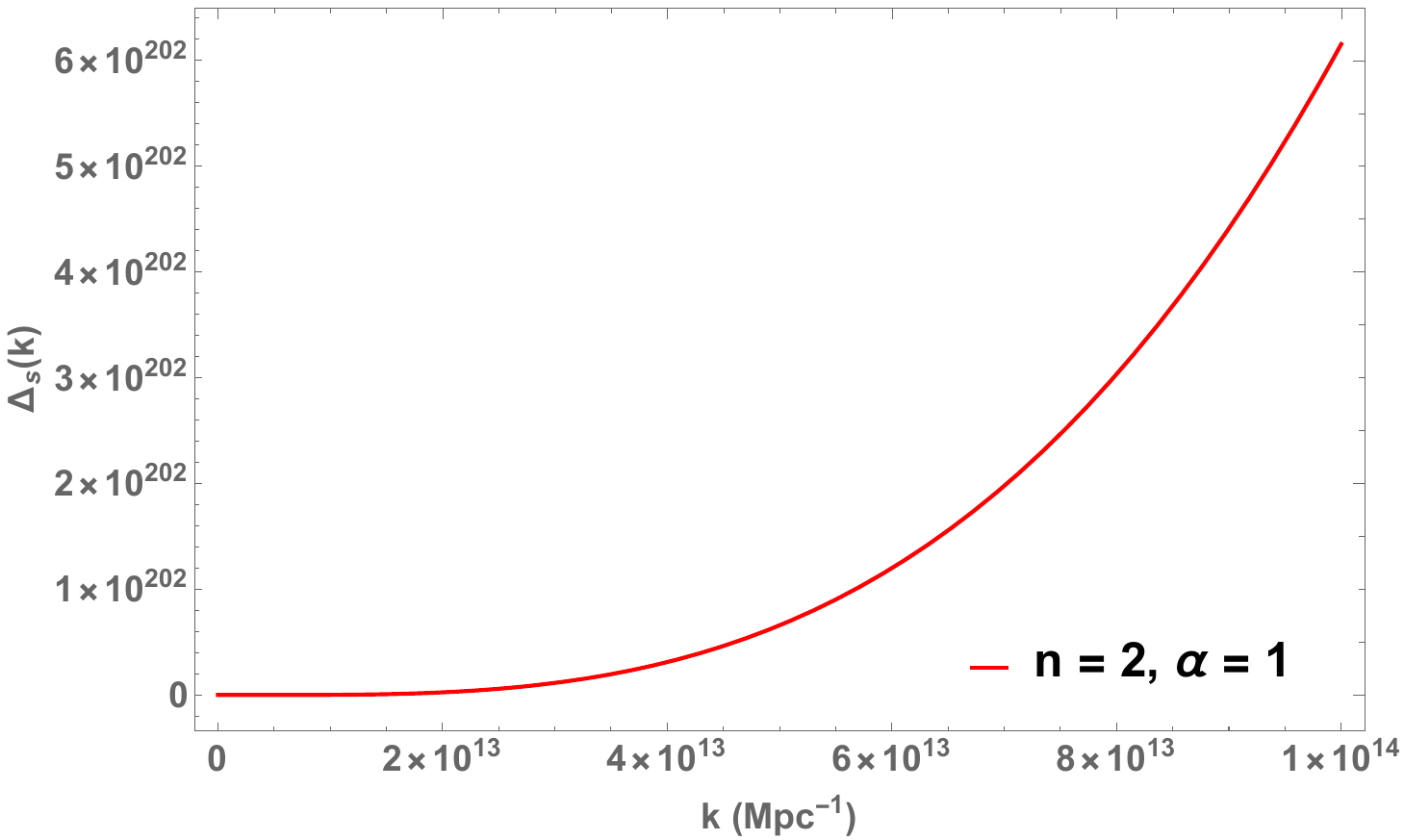"}
	\caption{The scalar power spectrum $\Delta_s(k)$ vs. $k$ in the same  high momentum range as in Figures  \ref{fig:potential-high-k} - \ref{fig:perturbed-inflaton-high-k}. Extremely high values of the power spectrum in this range indicate very high correlations among the mode momenta belonging to the quantized inflaton field, which is possible in the case of black hole formations.}
	\label{fig:Power_spectrum}
\end{figure}

 Figure \ref{fig:Power_spectrum} shows that, the scalar power spectrum, which is a measure of two-point correlations among the fluctuations, increases very rapidly at the $k$ values of the PBH formations. Such $k$-space behaviour of the scalar power spectrum signifies the breakdown of the perturbative framework, as well as a very high quantum correlation, which may be favourable for the PBH formations. The range of $k$, where $\Delta_s (k)$ shoots up in spatially flat gauge, is almost similar to that, derived from the solution of Mukhanov-Sasaki equation in co-moving gauge \cite{Mahbub:2019uhl}.

\par
Although the calculations from Figure \ref{fig:potential-high-k} through Figure \ref{fig:Power_spectrum} are with the chaotic $\alpha$-attractor $T$ model potential, the results will be similar for the corresponding $E$ model potential. Therefore, we are not displaying those results here.
\par
To further endorse the identification of the peaks of the density distribution over various $k$ modes (see Figure \ref{fig:Density_1}) with the PBHs formation events, in Table \ref{tab:table_name_parameters}  we have shown our results regarding some of their physical properties. \par
\begin{figure}[H]
    \centering
    \includegraphics[width=0.8\linewidth]{"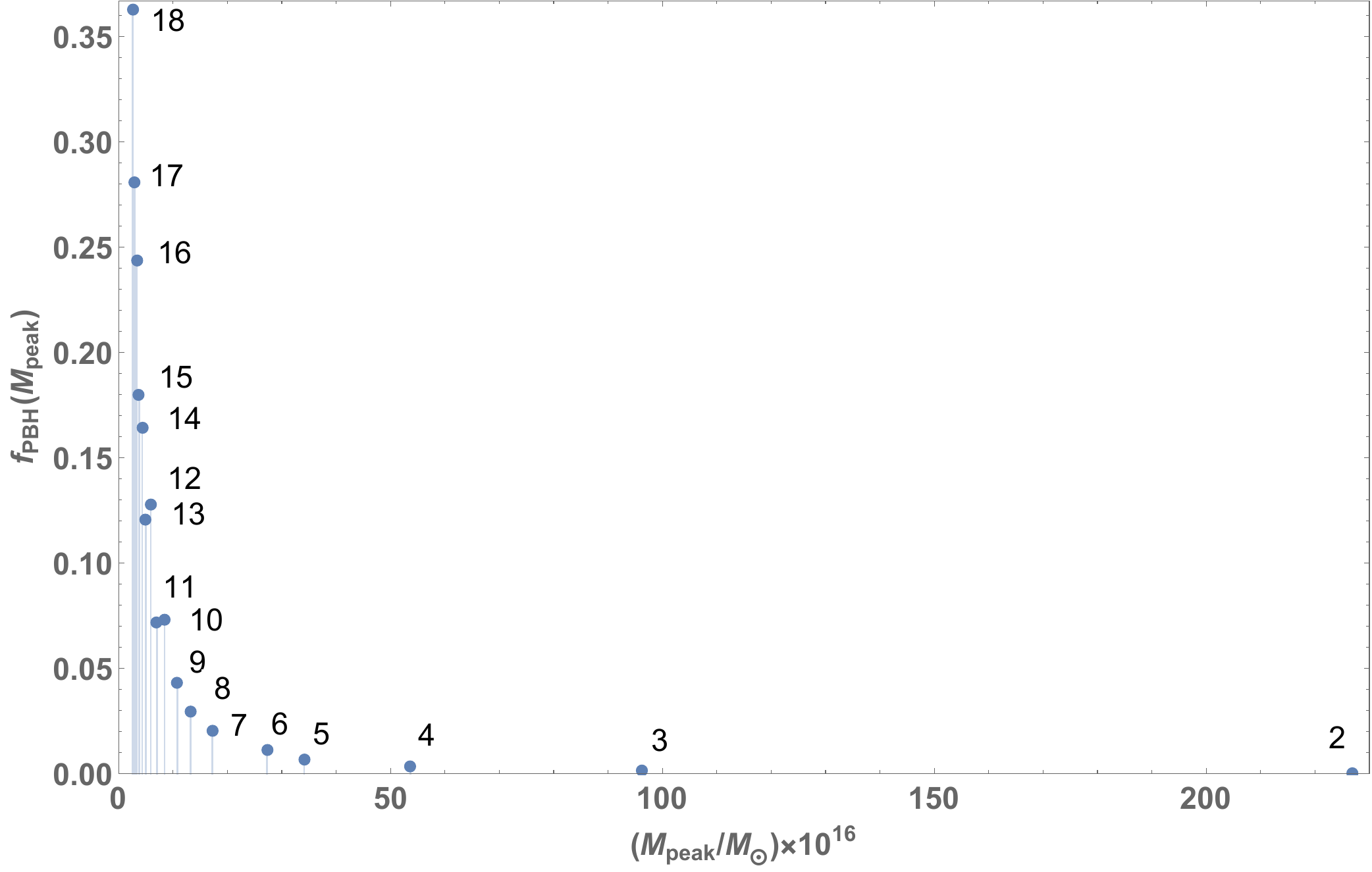"}
    \caption{ $f_\mathrm{PBH}(M_\mathrm{peak})$ vs. peak mass, $M_\mathrm{peak}$ (in the unit of $M_\odot$), of the PBHs corresponding to the enhancements in the density contrast profile in the $k$-space . The numbers with the blue dots indicate the peak-number for which the $f_\mathrm{PBH}$ is calculated. The value of $f_\mathrm{PBH}$ increases as the mass of the PBH decreases. Maximum obtained value of $f_\mathrm{PBH}$ is $0.36286$ for peak number $18$ (top-most dot) and the minimum value is $6.12\times10^{-6}$ for peak-number $1$ (not shown in the figure). Exact values can be seen in Table \ref{tab:table_name_fbh}.}
    \label{fig:fbh}
\end{figure}

\begin{figure}[H]
    \centering
    \includegraphics[width=0.8\linewidth]{"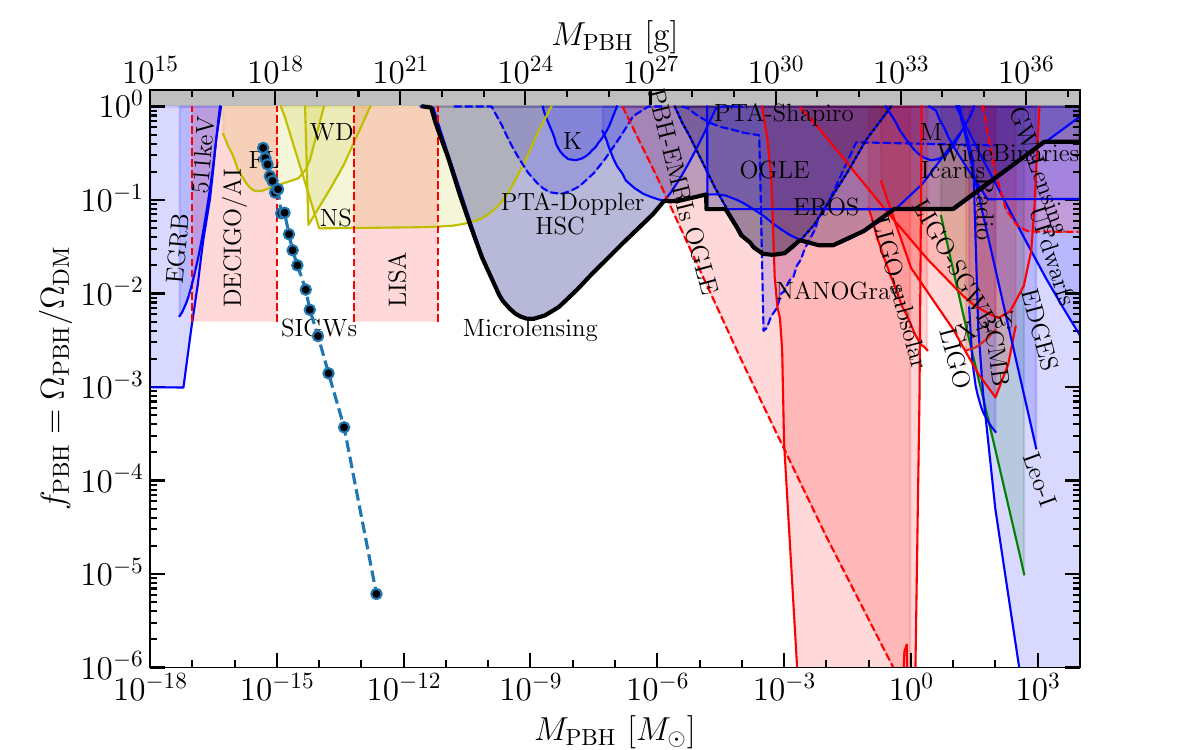"}
    \caption{Comparison of calculated PBH abundance in DM with that of the observational constraints in various experiments with respect to a monochromatic mass distribution. The black dots represent the data given in Table \ref{tab:table_name_fbh}, related to the $18$ peaks shown in Figure \ref{fig:Density_1}. It can be seen that the data corresponding to the 10th through 18th peak are consistent with the DECIGO/AI and  FL forecasts and that corresponding to the 4th through 9th peak overlap with SIGWs predictions. The results indicate that the PBHs found in our formalism are $\sim 0.3\%-36\%$ of the total DM present in the universe. Detailed information on the experiments can be found in \cite{bradley_j_kavanagh_2019_3538999}. } \label{fig:fbhpaper}
\end{figure}

 
In the column $4$ of Table \ref{tab:table_name_parameters}, the momentum-dependent PBH masses in the unit of the solar mass ($M_\odot$) have been presented. The calculation of the masses is based on the formula given in Eq. (\ref{eq:massPBH}). The mass-dependent evaporation times and the Hawking temperatures \cite{1975CMaPh..43..199H} from Eqs. (\ref{eq:evaporationtime}) and (\ref{eq:hawkingtemperture}) are presented in columns $5$ and $6$ respectively.  The evaporation time scale of the PBHs in our calculations is found in the range $10^{25}-10^{33}$s, which is very large in comparison to the age of the universe ($\sim 10^{17}$s). Also, the Hawking temperatures ($\sim 10^{-5} - 10^{-8}$ GeV) are very small, showing that the possibility of extinction of such PBHs by Hawking radiation is negligibly small. Thus, these PBHs may exist in the present universe, thereby making themselves substantial part of the DM.
\begin{table}[H]
\captionsetup{justification=centering,width=0.68\textwidth}
\caption{PBH properties corresponding to the peaks in Figure \ref{fig:Density_1}.}
\begin{center}
	\begin{adjustbox}{width=0.9\textwidth}
		\begin{tabular}{ |c|c|c|c|c|c| } 	
			\hline
		    Peak No. & $k_\mathrm{peak}$ (in Mpc$^{-1}$) & $(\frac{\delta\rho}{\rho^{(0)}})_\mathrm{rad}|_{k=k_\mathrm{peak}}$ & $M_\mathrm{PBH}$ (in $M_\odot$) & $t_\mathrm{eva}$ (in sec.) & $T_H$ (in GeV)\\
			\hline 
		1 & 0.43 $\times$ 10$^{13}$ & 30.78 & 1.35 $\times$ 10$^{-13}$ & 7.74 $\times$ 10$^{33}$ & 3.72 $\times$ 10$^{-8}$  \\	
		2 & 1.05 $\times$ 10$^{13}$ & 12.37 & 2.27 $\times$ 10$^{-14}$ & 3.67 $\times$ 10$^{31}$ & 2.21 $\times$ 10$^{-7}$  \\
		3 & 1.61 $\times$ 10$^{13}$ & 8.04 & 9.61 $\times$ 10$^{-15}$ & 2.79 $\times$ 10$^{30}$ & 5.23 $\times$ 10$^{-7}$  \\
		4 & 2.16 $\times$ 10$^{13}$ & 5.94 & 5.37 $\times$ 10$^{-15}$ & 4.86 $\times$ 10$^{29}$ & 9.36 $\times$ 10$^{-7}$  \\
		5 & 2.71 $\times$ 10$^{13}$ & 4.83 & 3.41 $\times$ 10$^{-15}$ & 1.24 $\times$ 10$^{29}$ & 1.47 $\times$ 10$^{-6}$  \\
		6 & 3.25 $\times$ 10$^{13}$ & 3.96 & 2.73 $\times$ 10$^{-15}$ & 4.18 $\times$ 10$^{28}$ & 2.12 $\times$ 10$^{-6}$  \\
		7 & 3.81 $\times$ 10$^{13}$ & 3.41 & 1.72 $\times$ 10$^{-15}$ & 1.60 $\times$ 10$^{28}$ & 2.92 $\times$ 10$^{-6}$  \\
		8 & 4.36 $\times$ 10$^{13}$ & 2.97 & 1.32 $\times$ 10$^{-15}$ & 7.23 $\times$ 10$^{27}$ & 3.81 $\times$ 10$^{-6}$  \\
		9 & 4.88 $\times$ 10$^{13}$ & 2.73 & 1.08 $\times$ 10$^{-15}$ & 4.05 $\times$ 10$^{27}$ & 4.62 $\times$ 10$^{-6}$  \\
		10 & 5.44 $\times$ 10$^{13}$ & 2.42 & 8.47 $\times$ 10$^{-16}$ & 1.91 $\times$ 10$^{27}$ & 5.93 $\times$ 10$^{-6}$  \\
		11 & 5.97 $\times$ 10$^{13}$ & 2.11 & 7.03 $\times$ 10$^{-16}$ & 1.09 $\times$ 10$^{27}$ & 7.15 $\times$ 10$^{-6}$  \\
		12 & 6.51 $\times$ 10$^{13}$ & 1.99 & 5.91 $\times$ 10$^{-16}$ & 6.54 $\times$ 10$^{26}$ & 8.49 $\times$ 10$^{-6}$  \\
		13 & 7.07 $\times$ 10$^{13}$ & 1.74 & 5.01 $\times$ 10$^{-16}$ & 3.95 $\times$ 10$^{26}$ & 1.00 $\times$ 10$^{-5}$  \\
		14 & 7.60 $\times$ 10$^{13}$ & 1.60 & 4.34 $\times$ 10$^{-16}$ & 2.57 $\times$ 10$^{26}$ & 1.15 $\times$ 10$^{-5}$  \\
		15 & 8.16 $\times$ 10$^{13}$ & 1.43 & 3.76 $\times$ 10$^{-16}$ & 1.66 $\times$ 10$^{26}$ & 1.33 $\times$ 10$^{-5}$  \\
		16 & 8.70 $\times$ 10$^{13}$ & 1.37 & 3.31 $\times$ 10$^{-16}$ & 1.14 $\times$ 10$^{26}$ & 1.51 $\times$ 10$^{-5}$  \\
		17 & 9.25 $\times$ 10$^{13}$ & 1.31 & 2.93 $\times$ 10$^{-16}$ & 7.89 $\times$ 10$^{25}$ & 1.72 $\times$ 10$^{-5}$ \\
		18 & 9.80 $\times$ 10$^{13}$ & 1.24 & 2.60 $\times$ 10$^{-16}$ & 5.53 $\times$ 10$^{25}$ & 1.93 $\times$ 10$^{-5}$ \\
		\hline
		\end{tabular}
		\end{adjustbox}
\end{center}
\label{tab:table_name_parameters}
\end{table}
In Table \ref{tab:table_name_fbh}, the statistical results of the the Eqs.(\ref{eq:sigma}),  (\ref{eq:rate}) and (\ref{eq:fpbh}) corresponding to the peaks of the density distribution in $k$ space, have been shown. We have calculated the mass scale dependent standard deviation $\sigma (M)$ (column $7$), formation rate $\beta(M)$ (column $8$) and the PBH abundance $f_\mathrm{PBH} (M)$ (column $9$) in total DM, around a peak mass $M_{\mathrm{peak}}$ at $k_{\mathrm{peak}}$ (given in Table \ref{tab:table_name_parameters}) within a specified $k$ range $k_{\mathrm{min}}-k_{\mathrm{max}}$ corresponding to $\Delta_{\mathrm{min}}-\Delta_{\mathrm{max}}$ above the threshold $\Delta_c$=$\Delta_{\mathrm{min}}$. The $\sigma (M)$ decreases from $7.681\times 10^{20}$ to $7.980\times 10^{15}$ as the heights of the peaks go down from $30.78$ to $1.24$. Therefore, the statistical error of the $k$ space mass distribution decreases as  height of the peaks becomes less. Therefore, the probability of formation of the corresponding PBHs should grow, which is indeed reflected in the increase of $\beta (M)$ from $1.577\times 10^{-20}$ to $4.099\times 10^{-17}$. Eventually the PBH abundance $f_\mathrm{PBH}(M)$, that is, the fraction of PBH present in DM is increased from $6.12\times 10^{-6}$ to $3.6286\times 10^{-1}$.
\begin{table}[H]
\captionsetup{justification=centering,width=0.68\textwidth}
\caption{Statistical properties of the PBHs related to the peaks in the density contrast profile of Figure \ref{fig:Density_1}. These properties are consistent with SIGWs, DECIGO/AI and FL forecasts.}
\begin{center}
	\begin{adjustbox}{width=0.9\textwidth}
		\begin{tabular}{ |c|c|c|c|c|c|c|c|c| } 	
			\hline
		    Peak No. & $k_\mathrm{min}$ (in Mpc$^{-1}$) & $k_\mathrm{max}$ (in Mpc$^{-1}$) & $\Delta_\mathrm{min}$ & $\Delta_\mathrm{max}$ & $M_\mathrm{peak}/M_\odot$ & $\sigma (M)$ & $\beta (M)$ & $f_\mathrm{PBH}(M)$\\
			\hline 
		1 & 1.12 $\times$ 10$^{12}$ & 7.09 $\times$ 10$^{12}$ & 0.42 & 30.78 & 1.35 $\times$ 10$^{-13}$ & 7.681 $\times$ 10$^{20}$ & 1.577 $\times$ 10$^{-20}$ & 6.12 $\times$ 10$^{-6}$ \\  
		2 & 8.28 $\times$ 10$^{12}$ & 1.30 $\times$ 10$^{13}$ & 0.42 & 12.37 & 2.27 $\times$ 10$^{-14}$ & 1.217 $\times$ 10$^{19}$ & 3.918 $\times$ 10$^{-19}$ & 0.000371 \\
		3 & 1.39 $\times$ 10$^{13}$ & 1.84 $\times$ 10$^{13}$ & 0.42 & 8.04 & 9.61 $\times$ 10$^{-15}$ & 3.167 $\times$ 10$^{18}$ & 9.597 $\times$ 10$^{-19}$ & 0.001397 \\ 
		4 & 1.95 $\times$ 10$^{13}$ & 2.38 $\times$ 10$^{13}$ & 0.42 & 5.94 & 5.37 $\times$ 10$^{-15}$ & 1.238 $\times$ 10$^{18}$ & 1.778 $\times$ 10$^{-18}$ & 0.00346 \\
		5 & 2.499 $\times$ 10$^{13}$ & 2.93 $\times$ 10$^{13}$ & 0.42 & 4.83 & 3.41 $\times$ 10$^{-15}$ & 6.263 $\times$ 10$^{17}$ & 2.809 $\times$ 10$^{-18}$ & 0.00686 \\
		6 & 3.05 $\times$ 10$^{13}$ & 3.45 $\times$ 10$^{13}$ & 0.42 & 3.96 & 2.73 $\times$ 10$^{-15}$ & 3.374 $\times$ 10$^{17}$ & 4.186 $\times$ 10$^{-18}$ & 0.01143 \\
		7 & 3.61 $\times$ 10$^{13}$ & 4.00 $\times$ 10$^{13}$ & 0.42 & 3.41 & 1.72 $\times$ 10$^{-15}$ & 2.046 $\times$ 10$^{17}$ & 5.829 $\times$ 10$^{-18}$ & 0.02006 \\
		8 & 4.15 $\times$ 10$^{13}$ & 4.54 $\times$ 10$^{13}$ & 0.42 & 2.97 & 1.32 $\times$ 10$^{-15}$ & 1.373 $\times$ 10$^{17}$ & 7.409 $\times$ 10$^{-18}$ & 0.02911 \\ 
		9 & 4.72 $\times$ 10$^{13}$ & 5.10 $\times$ 10$^{13}$ & 0.42 & 2.73 & 1.08 $\times$ 10$^{-15}$ & 9.263 $\times$ 10$^{16}$ & 9.948 $\times$ 10$^{-18}$ & 0.04320 \\
		10 & 5.30 $\times$ 10$^{13}$ & 5.60 $\times$ 10$^{13}$ & 0.42 & 2.42 & 8.47 $\times$ 10$^{-16}$ & 5.342 $\times$ 10$^{16}$ & 1.494 $\times$ 10$^{-17}$ & 0.07325 \\
		11 & 5.82 $\times$ 10$^{13}$ & 6.20 $\times$ 10$^{13}$ & 0.42 & 2.11 & 7.03 $\times$ 10$^{-16}$ & 5.047 $\times$ 10$^{16}$ & 1.336 $\times$ 10$^{-17}$ & 0.07190 \\
		12 & 6.40 $\times$ 10$^{13}$ & 6.68 $\times$ 10$^{13}$ & 0.42 & 1.99 & 5.91 $\times$ 10$^{-16}$ & 2.884 $\times$ 10$^{16}$ & 2.172 $\times$ 10$^{-17}$ & 0.12751 \\
		13 & 6.89 $\times$ 10$^{13}$ & 7.23 $\times$ 10$^{13}$ & 0.42 & 1.74 & 5.01 $\times$ 10$^{-16}$ & 2.784 $\times$ 10$^{16}$ & 1.891 $\times$ 10$^{-17}$ & 0.120604 \\
		14 & 7.46 $\times$ 10$^{13}$ & 7.76 $\times$ 10$^{13}$ & 0.42 & 1.60 & 4.34 $\times$ 10$^{-16}$ & 1.961 $\times$ 10$^{16}$ & 2.400 $\times$ 10$^{-17}$ & 0.16446 \\
		15 & 8.00 $\times$ 10$^{13}$ & 8.31 $\times$ 10$^{13}$ & 0.42 & 1.43 & 3.76 $\times$ 10$^{-16}$ & 1.647 $\times$ 10$^{16}$ & 2.447 $\times$ 10$^{-17}$ & 0.18012 \\
		16 & 8.57 $\times$ 10$^{13}$ & 8.85 $\times$ 10$^{13}$ & 0.42 & 1.37 & 3.31 $\times$ 10$^{-16}$ & 1.220 $\times$ 10$^{16}$ & 3.105 $\times$ 10$^{-17}$ & 0.24361 \\
		17 & 9.11 $\times$ 10$^{13}$ & 9.40 $\times$ 10$^{13}$ & 0.42 & 1.31 & 2.93 $\times$ 10$^{-16}$ & 1.054 $\times$ 10$^{16}$ & 3.369 $\times$ 10$^{-17}$ & 0.28098 \\
		18 & 9.66 $\times$ 10$^{13}$ & 9.92 $\times$ 10$^{13}$ & 0.42 & 1.24 & 2.60 $\times$ 10$^{-16}$ & 7.980 $\times$ 10$^{15}$ & 4.099 $\times$ 10$^{-17}$ & 0.36286 \\
		\hline
		\end{tabular}
		\end{adjustbox}
\end{center}
\label{tab:table_name_fbh}
\end{table}
In Figure \ref{fig:fbh} we have shown how the PBH abundance increases almost monotonically (except at peak number $10$ and $13$) with decrease of peak mass $M_\mathrm{peak}$, due to the fact stated above. This monotonous behaviour of $f_\mathrm{PBH} (M)$ might be a consequence of using monochromatic PBH mass distribution in $k$ space.
Figure \ref{fig:fbhpaper} depicts the comparison of our result (shown in Table \ref{tab:table_name_fbh} and Figure \ref{fig:fbh}) with experimental constraints. From this figure we see that PBHs corresponding to the small peaks and hence small masses are likely to be more favourable to constitute the DM in the present universe, so far as experimental forecasts are concerned. The heavy PBHs are hardly possible to form in $k$ space from the inflationary quantum fluctuations with the basic $\alpha$ attractor potentials, but they can be produced in sufficient amount when physical processes such as accretion or merger takes place \cite{Ali-Haimoud:2018dau,Raidal:2018bbj,Vaskonen:2019jpv}. However, in that case, instead of a monochromatic mass distribution, we would have to use an extended mass function \cite{Bellomo:2017zsr}. Nonetheless, an important observation associated with our results  is that, the signals from these small-mass PBHs will comprise gravitational waves of small frequencies. The DECIGO project is designed to detect such frequencies. The figure \ref{fig:fbhpaper} shows that the 10th to 18th peaks are consistent with the DECIGO/AI forecasts. These gravitational waves are referred to as induced gravitational waves \cite{Domenech:2021ztg}. SIGWs project provides the relevant constraints for the PBHs, producing such waves. Also, these PBHs are detected by a  special technique called Gamma Ray Bursts (GRBs)-femtolensing \cite{Katz:2018zrn}, used in the FL forecasts, which match with our results.

\section{Conclusions}
\label{sec:conclusions}
In conclusion, we have examined, in this paper, the possibility of the PBH formation in the high-momentum sub-horizon regime, where the inflationary perturbation breaks down. We have worked in the spatially flat gauge, where we can study the perturbations in the inflaton field as well as in the background gravitational field, and with the $\alpha$-attractor inflaton potentials. The present study may be summarized in the following points.
\begin{itemize}
\item[(i)] 
We have shown in Section \ref{sec:results} that, $\delta V (k)$ (Figure \ref{fig:potential-high-k}), $\delta\phi (k)$ (Figure \ref{fig:perturbed-inflaton-high-k}) and $\Delta_s (k)$ (Figure \ref{fig:Power_spectrum}) shoot up at $k\sim 10^{13}$ Mpc$^{-1}$ signalling the breakdown of linear perturbation. In the same range of $k$, density contrast $\Delta$ exceeds $\Delta_c$ (Figure \ref{fig:Density_1}) and Bardeen potential $\Phi_B (k)$ becomes large negative (Figure \ref{fig:Bardeen_1}), providing favourable condition for PBH formation. In Figure \ref{fig:pic2pic4} we have judged the efficacy of $\alpha$-attractor $T$ model potential as against the polynomial type $\phi^2$ potential in production of PBHs. In Table \ref{tab:table_name_parameters} we have seen that those PBHs have large evaporation time ($t_\mathrm{eva}=7.74\times 10^{3}-5.53\times 10^{25}$s) (see column $5$) and very small Hawking temperature ($T_H= 3.72\times 10^{-8}-1.93\times 10^{-5}$GeV) (see coloumn $6$), making them a potential constituent of dark matter. In Table \ref{tab:table_name_fbh} we have demonstrated the formation rate $\beta (M)$ ($1.577\times 10^{-20}-4.099\times 10^{-17}$) and the abundance $f_\mathrm{PBH} (M)$ ($6.12\times 10^{-6}-3.6\times 10^{-1}$) of those PBHs in total dark matter content. We have also observed that low mass PBHs are more favourable than heavy PBHs in constituting the DM as well as in probability of formation. Finally in Figure \ref{fig:fbhpaper} we have verified our results with the current observational bounds and we found that it matches with DECIGO/AI, FL and SIGW forecasts very well.\par

\item[(ii)] We have found the solutions of Eqs. (\ref{eq:17}), (\ref{eq:18}) and (\ref{eq:19}) as the domain of PBH formation in the $k$ range $10^{13}-10^{14}$ Mpc$^{-1}$ of peak-masses $1.35\times 10^{-13}-2.60\times 10^{-16}M_\odot$. The lower $k$ ($k<10^{13}$ Mpc$^{-1}$) solutions corresponding to heavy PBHs ($M_\mathrm{PBH}>10^{-13}M_\odot$) could not be obtained in our study because that would involve physical processes such as  accretion, merger and binary, which are outside the scope of the present paper. Therefore, our results do not match with the experimental forecasts beyond LISA (see Figure \ref{fig:fbhpaper}). 
\item[(iii)] One of the striking results, here, is that the Bardeen potential $\Phi_B (k)$ in the spatially flat gauge manifests as a driving force for accumulating mass around the inflaton perturbation, which leads to dynamical PBH formation in the radiation-dominated era when large $k$ sub-horizon modes re-enter the Hubble horizon in small conformal time.

\item[(iv)] In the spatially flat gauge, the inflaton potential itself remains unchanged in the $\phi$ space, which is also reflected from Figure \ref{fig:potential-high-k} where $V^{(0)} (k)$ does not change significantly in $k$ space. So, no amplification occurs in unperturbed part, which therefore does not contribute to the PBH formation. This essentially reflects the fact that PBH formation is not due to the original $\alpha$ attractor potentials in field space. On the other hand, the enhancement in the perturbed part $\delta V(k)$ (Figure \ref{fig:perturbed-potential-high-k}) is because of enhancement in $\delta \phi(k)$ (Figure \ref{fig:perturbed-inflaton-high-k}) and in this way the situation congenial for PBH formation occurs by the breakdown of perturbation and consequent enhancement in density contrast. In the formalism involving co-moving gauges, the potential involves inflection points \cite{Choudhury:2013woa, Ballesteros:2017fsr, Bhaumik:2019tvl} / ultra slow roll (USR) \cite{Mahbub:2019uhl,Mahbub:2021qeo,Biagetti:2021eep, 2021hllNg:}, which results in PBH formations. Thus, the final  result is independent of the gauge chosen: the comoving gauge or the spatially flat gauge which shows that it is gauge-independent or physical. The advantage of choosing the spatially flat gauge is that it is convenient for study of the inflation \cite{Baumann:2009ds} and also it takes into account of the perturbation in the inflaton field as well as in the metric.\par
Therefore, we conclude that, we have found the signatures of the PBH formation from the breakdown of linear cosmological perturbation in the gauge invariant way and the amplification of the chaotic $\alpha$ attractor $T$ model potential in $k$ space, without affecting its original form in the field space.

\end{itemize}


\acknowledgments
The present work has been carried out using some of the facilities provided by the University Grants Commission to the Center of Advanced Studies under the CAS-II program. CS and AS acknowledge the government of West Bengal for granting them the Swami Vivekananda fellowship. The Authors want to thank Dr. Bradley Kavanagh for useful correspondences regarding PBHbounds. BG(I) acknowledges the Department of Science and Technology for providing her the DST-Inspire Faculty Fellowship, and thanks Rajeev Kumar Jain, Nilanjandev Bhaumik and Jishnu Sai P for illuminating discussions.


\newpage
\bibliographystyle{utcaps}
\bibliography{biblio}

\end{document}